\PassOptionsToPackage{table}{xcolor}
\documentclass[sigconf]{acmart} 

\usepackage{subfigure}
\usepackage{amsmath,amsfonts}
\usepackage[algo2e,linesnumbered,ruled,lined,noend]{algorithm2e}
\SetKwInOut{Parameter}{Parameters}
\SetKwInOut{Constraint}{Constraints}
\SetKwFunction{Mining}{Mining}
\SetKwFunction{MiningTwo}{Mining2}
\SetKwFunction{Candidates}{Candidates}
\SetKwProg{Fn}{Function}{}{}
\usepackage{graphicx}
\usepackage{textcomp}
\usepackage{color}
\usepackage{xcolor}
\usepackage{xspace}
\usepackage{soul}

\def\BibTeX{{\rm B\kern-.05em{\sc i\kern-.025em b}\kern-.08em
    T\kern-.1667em\lower.7ex\hbox{E}\kern-.125emX}}
    
\usepackage{multirow} 
\usepackage{booktabs} 

\usepackage[normalem]{ulem}
\useunder{\uline}{\ul}{}

\usepackage{pgf}
\usepackage{enumitem}
\setlist[itemize]{leftmargin=*}
\usepackage{stfloats}

\usepackage{makecell}

\usepackage{bbold}
\usepackage{mathtools}
\usepackage{comment}
\usepackage{microtype}
\usepackage[misc]{ifsym}
\usepackage{url}
\SetCommentSty{mycommfont}

\newcommand{\red}{\textcolor{red}}

\definecolor{best}{RGB}{120,220,190}

\definecolor{lightgreen}{rgb}{0.88, 1, 0.88}
\definecolor{openaigreen}{HTML}{10A37F}

\usepackage{fontawesome5}

\newcommand{\best}[1]{\cellcolor{blue!15}\textbf{#1}}
\newcommand{\bestbox}[1]{\colorbox{blue!15}{#1}}

\definecolor{rowgray}{gray}{0.92} 

\newcommand{\method}{{IIRG}\xspace}

\newcommand{\cmark}{\textcolor{green!60!black}{\ding{51}}}
\newcommand{\xmark}{\textcolor{red!80!black}{\ding{55}}}

\definecolor{NextItemBg}{HTML}{F6E4D8}
\definecolor{CollabBg}{HTML}{DEF1D3}
\definecolor{SemanticBg}{HTML}{D2EDFA}

\DeclareRobustCommand{\nextbg}[1]{{\sethlcolor{NextItemBg}\hl{#1}}}
\DeclareRobustCommand{\collabbg}[1]{{\sethlcolor{CollabBg}\hl{#1}}}
\DeclareRobustCommand{\semanticbg}[1]{{\sethlcolor{SemanticBg}\hl{#1}}}

\usepackage{multirow}
\usepackage{multicol}
\usepackage{pifont}
\usepackage[most]{tcolorbox}

\newtcolorbox{promptblock}{
  enhanced,
  colback=gray!3,
  colframe=black!45,
  boxrule=0.5pt,
  arc=2pt,
  left=7pt,
  right=7pt,
  top=7pt,
  bottom=7pt
}

\AtBeginDocument{%
  \providecommand\BibTeX{{%
    \normalfont B\kern-0.5em{\scshape i\kern-0.25em b}\kern-0.8em\TeX}}}

\copyrightyear{2027}
\acmYear{2027}
\acmConference[WWW '27]{Proceedings of the ACM Web Conference 2027}{May 10--14, 2027}{Dublin, Ireland}

\setcopyright{none}

\ccsdesc[500]{Information systems~Recommender systems}

\keywords{large language model, generative recommendation}

\setcopyright{none}
\renewcommand\footnotetextcopyrightpermission[1]{}

\begin{document}


    \title[Memorization in LLM-Based Generative Recommendation: Observations, Implications, and Improved Training]{Memorization in LLM-Based Generative Recommendation: Observations, Implications, and Improved Training}
    
    \settopmatter{authorsperrow=4}
	
    \author{Sunwoo Kim}
\authornote{Work done while at Snap Inc.}
\affiliation{%
    \institution{KAIST}
    \city{Seoul}
    \country{Republic of Korea}
}
\email{kswoo97@kaist.ac.kr}

\author{Sunkyung Lee}
\authornotemark[1]
\affiliation{%
    \institution{Sungkyunkwan University}
    \city{Suwon}
    \country{Republic of Korea}
}
\email{sk1027@skku.edu}

\author{Clark Mingxuan Ju}
\affiliation{%
    \institution{Scaled Cognition}
    \city{Bellevue}
    \state{WA}
    \country{USA}
}
\email{mju2@alumni.nd.edu}

\author{Donald Loveland}
\affiliation{%
    \institution{Snap Inc.}
    \city{Bellevue}
    \state{WA}
    \country{USA}
}
\email{dloveland@snapchat.com}

\author{Bhuvesh Kumar}
\affiliation{%
    \institution{Snap Inc.}
    \city{Bellevue}
    \state{WA}
    \country{USA}
}
\email{bkumar4@snapchat.com}

\author{Kijung Shin}
\affiliation{%
    \institution{KAIST}
    \city{Seoul}
    \country{Republic of Korea}
}
\email{kijungs@kaist.ac.kr}

\author{Neil Shah}
\affiliation{%
    \institution{Pinterest}
    \city{Seattle}
    \state{WA}
    \country{USA}
}
\email{neilshah@pinterest.com}

\author{Liam Collins}
\affiliation{%
    \institution{Snap Inc.}
    \city{Bellevue}
    \state{WA}
    \country{USA}
}
\email{lcollins2@snap.com}

    \begin{abstract}

Generative recommendation (GR) has emerged as a promising direction for recommender systems. 
Recently, large language models (LLMs) have been increasingly adopted for GR, as their rich pretrained knowledge is expected to help them generalize beyond common user behavior patterns that traditional memorization-oriented baselines can capture.
However, existing LLM-based GR works largely ignore LLMs' well-known {\em tendency to memorize} \cite{satvaty2024undesirable,li2025skewed}, which, if present in LLMs fine-tuned for GR, may affect their utilization of pretrained knowledge. 
In this work, we investigate this possibility by examining \textit{one-hop memorization}, where a model recommends items that are \textit{direct successors} of items in the training data. 
We show that LLMs rely more heavily on one-hop memorization than TIGER, a representative non-LLM-based GR model, and generally achieve larger relative gains for users whose target items are covered by such memorization.
We intuit that improving performance on the remaining users requires LLMs to learn richer item--item relations that complement one-hop transitions.
To address this, we propose \textbf{\method}, an analysis-driven training strategy that jointly trains LLMs to predict next items and generate \textit{multi-hop collaborative} and \textit{semantic} neighbors. 
This item-to-item supervision provides complementary relations beyond training-observed one-hop transitions.
We show that \method improves over LLMs trained solely with standard next-item prediction, with especially large gains for users whose test items are not covered by train-time one-hop transitions.
    \end{abstract}
	
	\maketitle

    
    \section{Introduction}
    \label{sec:introduction}

Generative recommendation (GR) has recently drawn growing interest within the recommendation community, particularly in the Web research community~\cite{wang2025generative, xie2025lohrec, wang2026mtgenrec}.
Given a user’s item interaction history, GR methods use a generative model to generate the identifier of the target item, rather than scoring a predefined set of candidate items~\cite{liu2025onerec, zheng2026dikgrec, qu2026diffusion}.

\begin{figure}[t]
\centering
    \includegraphics[width=\linewidth]{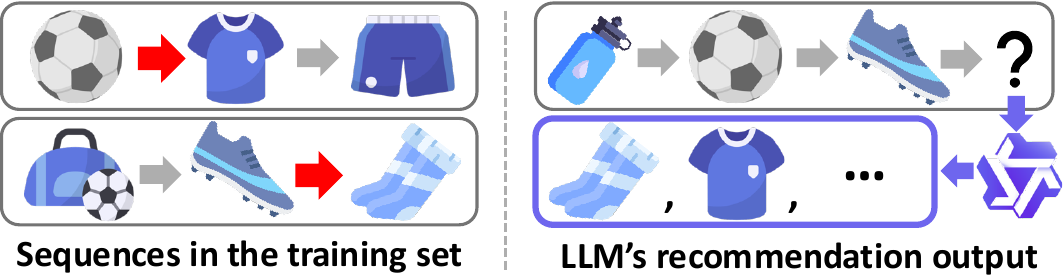}
    \caption{
    \textit{An example of one-hop memorization.}
    The LLM memorizes that soccer shirts and socks are purchased after soccer balls and cleats, respectively (\red{red} arrows), then recommends the shirts and socks to users who bought the balls and cleats.}
    \label{fig:memorization_example}
\end{figure}

A growing line of work on GR adopts large language models (LLMs) as recommender systems, motivated by their rich pretrained world knowledge~\cite{zhou2025openonerec, xie2026agentictagger, lin2025order, lin2026token, lin2026verifiable}.
These approaches have achieved strong performance on recommendation tasks in both academic~\cite{lee2025gram,liu2025understanding, zhu2026iterative} and industrial~\cite{he2026plum, d2026deploying} settings.

For GR, LLMs are often fine-tuned using next-item prediction because pretrained LLMs are not aware of the specific item set from which they are allowed to recommend, nor do they have sufficient knowledge of user--item interaction patterns.
Specifically, given the earlier part of a user’s item interaction sequence, LLMs are fine-tuned to generate the identifier (e.g., semantic ID or term ID) of the next item in the sequence~\cite{hong2025eager}. 

While LLMs hold promise for GR through their pretrained knowledge, prior work in NLP suggests a potential obstacle to fulfilling this promise. 
LLMs are well-known to \textit{memorize} (i.e., retain certain portions of the training set and later reproduce them at inference time~\cite{satvaty2024undesirable, li2025skewed}). 
Heavy reliance on such behavior may limit the ability of LLM-based GR models to capitalize on pretrained world knowledge, as their predictions would instead be driven by repetitions of patterns observed in the recommendation training data, just like traditional, non-language-informed recommender systems. 
However, this phenomenon remains largely underexplored in LLM-based GR.

To address this gap, we analyze memorization behavior in LLM-based GR.
We focus on a widely used LLM-based GR pipeline in which each item is represented by its text features alongside a structured item ID (e.g., semantic ID (SID) or term ID (TID)), which may be trained to align with the LLM's vocabulary \cite{he2026plum}. 
The LLM is then fine-tuned to generate the ID of the next item given the user’s previous interaction sequence~\cite{lee2025gram, zhang2026unleashing}. 
In this setting, we ask two questions: 
(1) \textit{what distinctive memorization behavior do LLMs exhibit compared to non-LLM-based GR models}, and 
(2) \textit{how does this behavior relate to recommendation performance?}

\begin{figure}[t]
\centering
    \includegraphics[width=\linewidth]{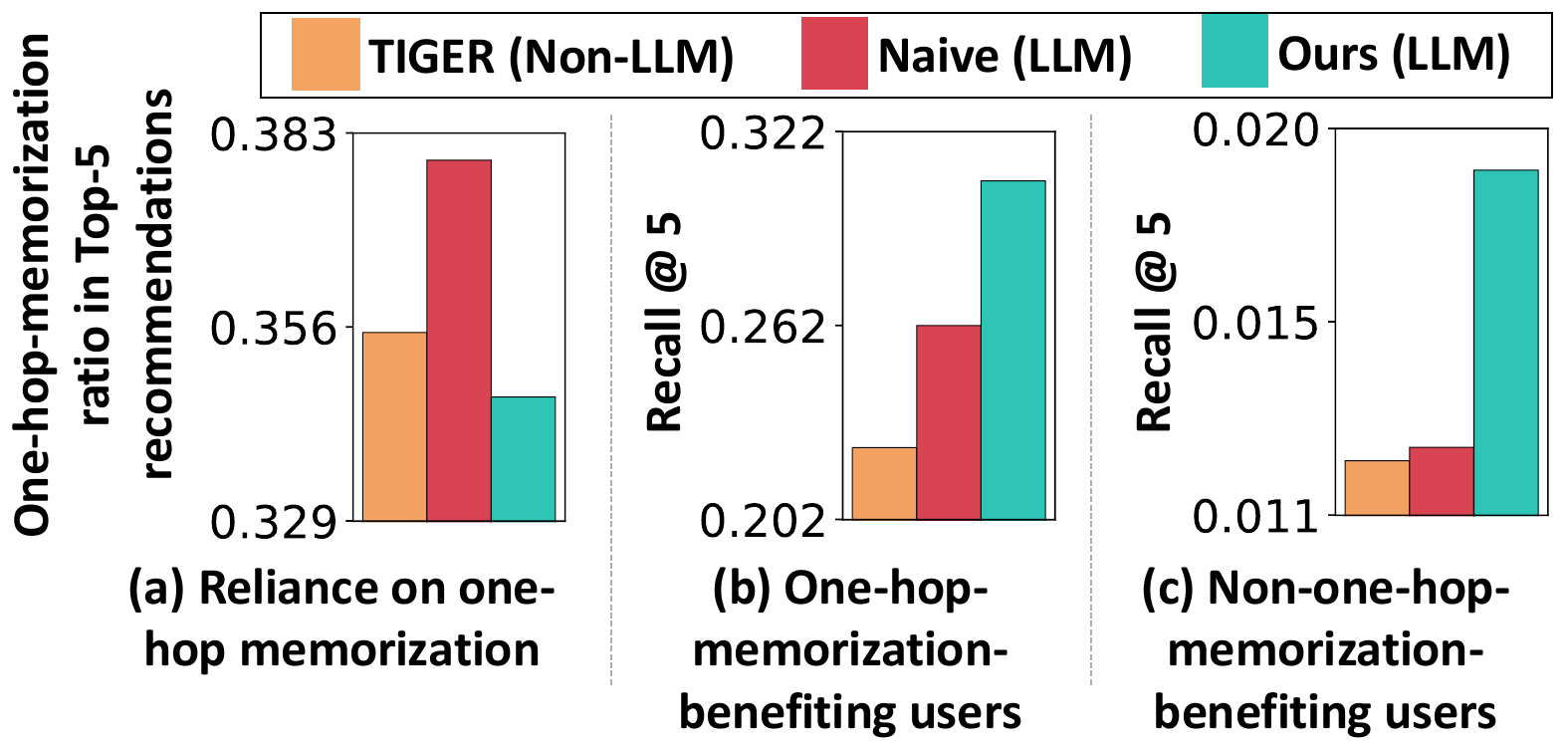}
    \caption{
    \textit{One-hop memorization in Sports.}
    LLMs using SIDs trained solely with next-item prediction (Naive) rely more on one-hop memorization than TIGER. 
    This reliance leads to gains concentrated among users covered by such memorization, with limited benefits for the remaining users. 
    \method, our training method, reduces this reliance and yields stronger performance gains for the remaining users.
    Details are in \S~\ref{sec:analysis}.
    }
    \label{fig:main_motivation}
\end{figure}

For the first question, we show that LLMs exhibit strong \textit{one-hop memorization}. 
In particular, LLMs memorize one-hop item transitions in the training set—transitions between consecutive items in user sequences—and at inference time base their recommendations on the observed successors of items in the user’s input sequence, as illustrated in Figure~\ref{fig:memorization_example}.
This memorization behavior is more pronounced in LLMs than in TIGER~\cite{rajput2023recommender}, a representative non-LLM-based GR model that uses a randomly initialized T5~\cite{raffel2020exploring} backbone, as shown in Figure~\ref{fig:main_motivation}(a).
This phenomenon persists after scaling TIGER to a near-LLM size and applying regularization to LLMs, suggesting that model size and overfitting alone are insufficient explanations.

For the second question, we show that LLMs often achieve larger relative performance gains over TIGER for users who benefit from one-hop memorization than for those who do not.
Specifically, we divide users into two groups: (1) users whose target item appears in the training set as a next item of any item from their input sequence, and (2) users whose target item does not.
We find that LLMs generally achieve larger relative gains over TIGER for the first group than for the second group; see Figure~\ref{fig:main_motivation}(b) and (c).

Given that one-hop memorization provides limited information for the target item for users in the second group, we hypothesize that improving performance for these users requires LLMs to learn richer item--item relations beyond one-hop transitions—relations that can connect the user’s input items to the target item.
To achieve this goal, we propose \textbf{\method} 
(\textbf{\underline{I}}tem–\textbf{\underline{I}}tem 
\textbf{\underline{R}}elation 
\textbf{\underline{G}}eneration), 
an analysis-driven training strategy that jointly trains LLMs on the next-item prediction task and two explicit item-to-item generation tasks: (1) collaborative-neighbor generation and (2) semantic-neighbor generation.

For collaborative relations, \method uses items that frequently co-occur with the given item across multiple hops in user interaction sequences.
For semantic relations, \method uses items with semantically relevant textual descriptions, helping the model capture items users may naturally consider together (e.g., related LEGO products or compatible accessories).
Our data analysis shows that these relations benefit users outside the coverage of one-hop memorization, reaching a group comparable in size to the users benefiting from one-hop memorization.
Note that \method is (1) identifier-agnostic, supporting both SIDs and TIDs, and (2) applicable across diverse LLM families (e.g., Qwen~\cite{yang2025qwen3}, LLaMA~\cite{grattafiori2024llama}, and Phi~\cite{abouelenin2025phi}).

{Notably, the key novelty of \method is twofold. 
At the task level, it introduces item-neighbor generation as auxiliary supervision for LLM-based generative recommendation. 
At the training-design level, it autoregressively generates relevance-ordered neighbor sequences rather than predicting neighbors independently.}\footnote{{The effectiveness of our task relative to existing auxiliary training tasks is demonstrated in Section~\ref{subsec:mainperformance}. Comparisons with alternative training designs, including shuffled-order generation and independent item–neighbor prediction, are provided in Appendix~\ref{subapp:alternativetraining}; all these alternatives underperform our proposed design.}}

Through experiments, we show that LLMs trained with \method{} consistently improve over an LLM trained solely with next-item prediction, achieving a 21\% average gain in Recall@5.
The gain is especially pronounced for users whose target items are not covered by one-hop memorization, yielding a 50\% gain compared to a 17\% gain for users whose target items are covered by such memorization in the Sports dataset (see Figure~\ref{fig:main_motivation}).
Our key contributions are summarized as follows:
\begin{itemize}[leftmargin=*]
    \item \textbf{\textit{C1. Behavioral analysis:}} We identify one-hop memorization as a prominent behavior of pretrained LLM-based GR and show that its relationship with recommendation performance differs substantially across users.
    \item \textbf{\textit{C2. Item-to-item generation:}} We propose \method{}, which jointly trains LLMs on next-item prediction and explicit generation of multi-hop collaborative and semantic neighbors.
    \item \textbf{\textit{C3. Design-goal validation:}} We show that \method reduces one-hop memorization and consistently improves recommendation performance across item IDs and LLM families, especially for users outside one-hop coverage.
\end{itemize}
For reproducibility, we provide our code and datasets in~\cite{iirg_anonymous_github}.

    
    \section{Related work}
    \label{sec:relatedwork}
    

\noindent\textbf{Memorization in LLMs and recommender systems.}
LLMs are known to reproduce exact phrases from their pre-training~\cite{kassem2023preserving} and/or fine-tuning data~\cite{mireshghallah2022empirical}, though the exact causes remain unclear. 
Hypothesized explanations involve factors such as model size or duplicated training data~\cite{satvaty2024undesirable}.
In many LLM applications, this reproduction can leak private or copyrighted material~\cite{henderson2023foundation}. Several methods have been studied to address this (e.g., data deduplication~\cite{lee2022deduplicating} and specialized decoding~\cite{ippolito2023preventing}).
In the recommendation literature, \citet{ding2026well} compared memorization in generative and ID-based recommender systems, but neither examined LLMs nor analyzed the overlap between top-ranked recommendations and memorization-based recommendations, both of which are central to our work.\footnote{Appendix~\ref{subapp:unifiedanalysis} provides a unified analysis covering all recommender systems and memorization metrics examined in either our work or~\citet{ding2026well}.
The results of this unified analysis support the same key conclusion as our analysis in \S~\ref{sec:analysis}.}
Recently, \citet{di2025llms} reported that many LLMs possess prior knowledge of movies in the MovieLens-1M dataset~\cite{harper2015movielens}, likely acquired during their pretraining. 
However, LLM memorization arising from post-training (i.e., fine-tuning for recommendation) remains underexplored and is the focus of our work.

\noindent\textbf{Generative recommendation with LLMs.}
Generative recommendation (GR) has been extensively studied in both academia and industry~\cite{rajput2023recommender, li2024large, xie2025lohrec, lin2025efficient, yang2025earn, penha2025semantic, wang2025generative}.
Among these, a promising direction leverages pretrained LLMs as the backbone for GR to exploit their rich world knowledge~\cite{penha2025semantic}.
Many approaches use structured item identifiers to mitigate hallucination and facilitate efficient decoding, most notably 
{semantic IDs} (SIDs)~\cite{kong2025minionerec, he2026plum} and {term IDs} (TIDs)~\cite{tan2024idgenrec,lee2025gram}.
After associating each item with such an ID, the LLM is fine-tuned for next-item prediction via autoregressive training on users' past item interaction sequences~\cite{zhang2026unleashing,he2026plum}.

\noindent\textbf{Auxiliary training tasks for LLM-based GR.}
To improve the GR performance of LLMs, several auxiliary training tasks have been leveraged, often alongside next-item prediction.\footnote{Here, auxiliary tasks refer to tasks used for LLM training, rather than those used for ID generation, such as the contrastive learning objective in PLUM~\cite{he2026plum}.}
These tasks can be categorized into three types:
(1) distillation tasks, which help lightweight LLM-based GR models acquire the reasoning abilities of stronger LLMs~\cite{hong2025eager}; (2) alignment tasks, which help LLMs better understand the identifiers of items~\cite{zhang2026unleashing}; (3) auxiliary recommendation tasks, such as rating prediction~\cite{geng2022recommendation} or item unmasking from user sequences~\cite{cao2024aligning}.
Note that, unlike these methods, \method trains LLMs to generate multi-hop collaborative and semantic neighbors, motivated by our one-hop memorization analysis.

    \section{Analysis of Memorization in GR}
    \label{sec:analysis}
    
In this section, we first formalize the generative recommendation (GR) setting. 
Then, we analyze the one-hop memorization behavior (i.e., a model’s reliance on one-hop transitions observed in the recommendation training data) of LLMs in comparison with non-LLM-based GR models.

\subsection{Problem Formulation}
\label{subsec:preliminary}
Let $\mathcal{U}$ and $\mathcal{I}$ denote the user and item sets, respectively.
Each user $u \in \mathcal{U}$ is associated with an item interaction sequence $s_u=(i^{(u)}_{1},i^{(u)}_{2},\cdots,i^{(u)}_{\vert s_{u} \vert})$,~\footnote{While $\mathcal{\vert A\vert}$ typically denotes the cardinality of a set $\mathcal{A}$, we use $\vert \cdot\vert$ to denote the length of a sequence unless otherwise stated, as our notation primarily concerns sequences.}
and each item $i_k \in \mathcal{I}$ is represented by its textual metadata \texttt{t}$(i_k)$, such as its title and description, and its identifier \texttt{id}$(i_k)$, such as a semantic ID (SID)~\cite{he2026plum} or term ID (TID)~\cite{zhang2026unleashing}.
Given a user interaction sequence $s_u$, the goal of generative recommendation (GR) is to generate the identifier of the item that user $u$ is most likely to interact with next.

\subsection{Analysis setup}\label{subsec:analysis-setup}

\noindent\textbf{Overview.}
We consider \textit{one-hop memorization} as a model’s inference-time reliance on one-hop item transitions observed during recommendation training—that is, its tendency to recommend training-observed successors of items in a test user’s interaction sequence.\footnote{We regard this as a behavior-level form of memorization; see Appendix~\ref{subapp:memorizationscope} for details.}
To quantify one-hop memorization, we measure the fraction of each model’s top-5 recommendations that correspond to memorized successors of items in the test user’s input sequence.
Then, we examine how the performance gains of LLM-based GR models over non-LLM-based GR models are related to one-hop memorization, motivating improvements in LLM-based GR models.

\noindent\textbf{Models and datasets.} 
We consider two GR models: 
(1) our target model, a Qwen-3.5-4B~\cite{yang2025qwen3} LLM trained with a next-item prediction objective under SIDs and TIDs (see Appendix~\ref{subapp:ourtermid} for how the IDs are constructed); and 
(2) TIGER~\cite{rajput2023recommender}, a representative non-LLM-based GR baseline.
We use three Amazon Reviews datasets~\cite{he2016ups}, Sports, Toys, and Beauty. 

\begin{figure}[t]
    \centering
    \includegraphics[width=0.9\linewidth]{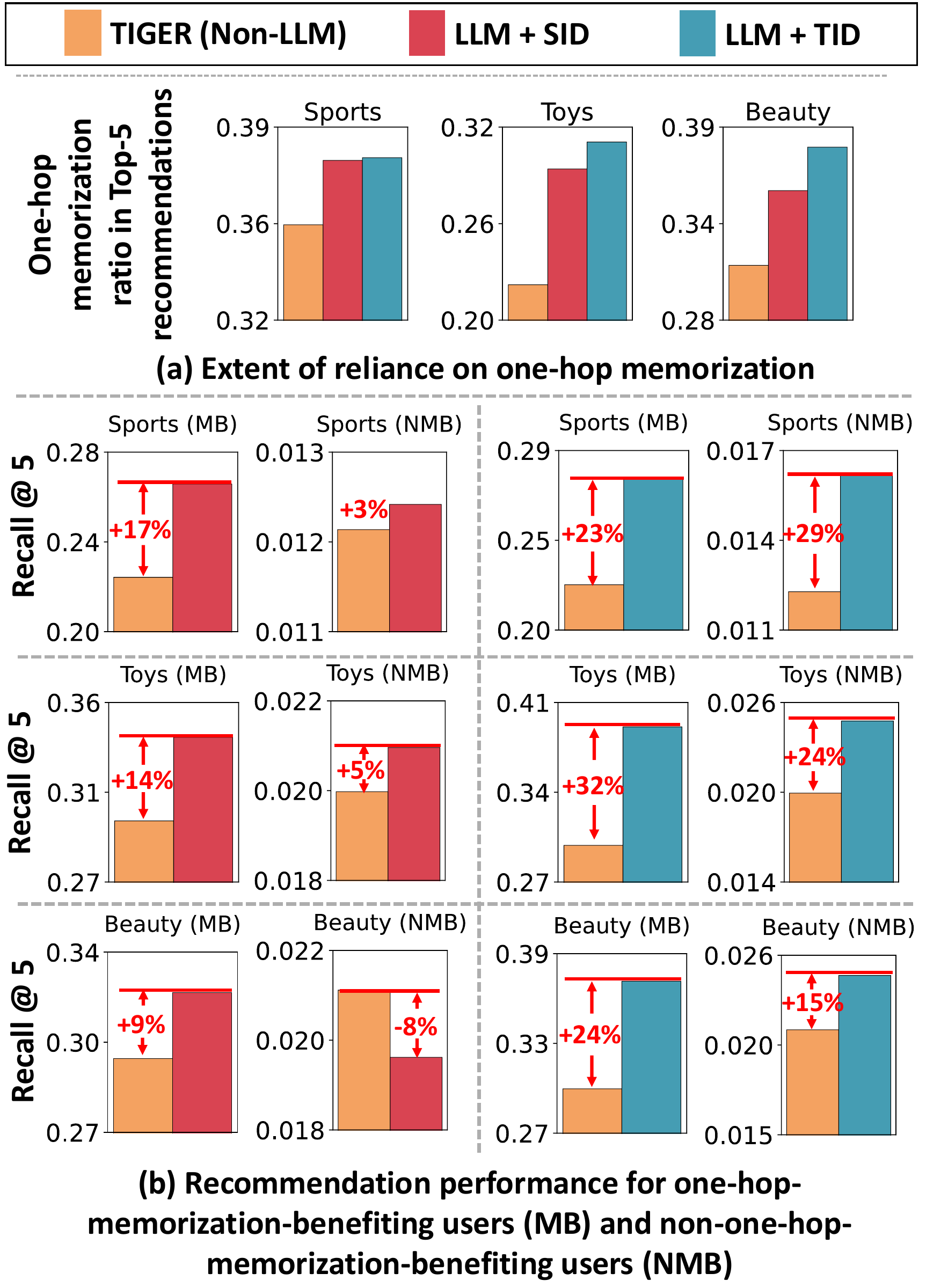}
    \caption{
    \textit{One-hop memorization and recommendation performance.}
    LLMs trained solely with next-item prediction exhibit stronger one-hop memorization than TIGER, a non-LLM-based model (see (a)). 
    The relative improvement of LLMs over TIGER is larger for one-hop-memorization-benefiting users than for non-one-hop-memorization-benefiting users in 5 out of 6 settings (see (b)).
    }
    \label{fig:main_observation}    
\end{figure}

\noindent\textbf{Training.}
Following~\citet{zhang2026unleashing}, the LLM is trained for next-item prediction using next-token prediction for a language-formatted item interaction sequence.
For the sequence representation, we represent each item $i_{k}$ in user $u$'s interaction sequence $s_{u}$ by its
identifier \texttt{id}$(i_{k})$ (either an SID or TID) and title \texttt{t}$(i_{k})$.
Appendix~\ref{subapp:next-item-prompt} provides example prompts for next-item prediction.
Based on this input structure, we perform supervised fine-tuning of the LLM.
Formally, let $\mathcal{T}^{(s_{u})}$ denote the tokenized representation of sequence $s_u$.
Then, the LLM training loss for user $u$ is:
\begin{equation}\label{eq:nextitempredloss}
\mathcal{L}_{u}
=
-\sum_{t=T+1}^{\left|\mathcal{T}^{(s_u)}\right|}
\frac{\log p_{\theta}
\left(
\mathcal{T}^{(s_u)}_{t}
\mid
\mathcal{T}^{(s_u)}_{< t}, \mathcal{C}
\right)
}
{\vert \mathcal{T}^{(s_u)}\vert - T},
\end{equation}
where $p_{\theta}$ denotes the token-generation probability of the LLM parameterized by $\theta$, $\mathcal{T}^{(s_u)}_{t}$ is the $t$-th token in sequence $\mathcal{T}^{(s_u)}$, $T$ is the index of the final token for the anchor item $i^{(u)}_{1}$ within $s_{u}$, and $\mathcal{C}$ denotes the tokens representing the instruction of next-item prediction.
We train the baseline non-LLM-based GR model, TIGER, following its original paper~\citep{rajput2023recommender}.

\noindent\textbf{Metrics.}
To quantify a model’s reliance on one-hop memorization, we define the adjacent neighbor set $\mathcal{N}(i_{k})$ as the multiset of all items that immediately follow $i_{k}$ in any training user sequence.
Next, we define a \textit{one-hop-based candidate set} for each user. 
This set is constructed by: 
(1) aggregating the adjacent neighbors of all items in the user's sequence (i.e., $\mathcal{M}_u = \uplus_{i_k \in s_u} \mathcal{N}(i_k)$ where $\uplus$ denotes the additive union of multisets) and 
(2) selecting the $\min(50,\vert \bigcup_{i_{k}\in s_{u}} \mathcal{N}(i_{k})\vert)$ items with largest multiplicities in $\mathcal{M}_u$ and tie-breaking based on popularity.\footnote{We use the top-$K$ most frequent adjacent neighbors to avoid excessively large candidate sets, which provide limited information for measuring one-hop memorization.}
Lastly, for each model, we compute the average fraction of its top-$5$ recommendations that belong to the one-hop-based candidate set.
In Appendix~\ref{subapp:cutoff}, we provide additional analyses with candidate sets constructed using cutoff thresholds other than $50$, as well as without any cutoff, and show that our findings remain consistent across these settings.

\noindent\textbf{User groups.}
To further investigate the impact of one-hop memorization on recommendation performance,
we divide users into two groups: the \textit{one-hop-memorization-benefiting} group and the \textit{non-one-hop-memorization-benefiting} group.\footnote{Here, \textit{benefiting} denotes potential target coverage by one-hop transitions rather than a causal estimate of realized benefit.} 
The former consists of users whose one-hop candidate set contains their target item, while the latter consists of the rest. 
We then measure the recommendation performance of each GR model on both groups to identify where the performance gains of LLMs primarily come from.

\subsection{Analysis results}\label{subsec:analysis-result}

\noindent\textbf{Reliance on one-hop memorization.}
As shown in Figure~\ref{fig:main_observation}(a), LLM-based GR models consistently exhibit a higher one-hop memorization ratio in their top-5 recommendations than TIGER, the non-LLM-based GR baseline. 
This result suggests that LLMs rely more heavily on one-hop memorization than the non-LLM-based GR model.
Notably, this finding holds for both TIDs and SIDs, indicating that this behavior is not specific to a particular type of item ID.
Moreover, this tendency remains even when we apply regularization techniques to LLMs, as detailed in Appendix~\ref{subapp:regularization}.
This result implies that overfitting alone is insufficient to explain the result.

\noindent\textbf{User-wise performance.}
As shown in Figure~\ref{fig:main_observation}(b), LLMs achieve larger relative performance improvements over TIGER for one-hop-memorization-benefiting users than for non-memorization-benefiting users in 5 out of 6 settings. 
This result implies that LLMs’ reliance on one-hop memorization primarily benefits users whose target items are covered by memorized one-hop transitions, but provides limited benefits for the rest of the users.
Notably, the overall trend persists when we control for test-item popularity by grouping users whose test items have similar popularity.

\noindent\textbf{Key implications.}
Our analysis suggests that relying on one-hop memorization offers limited benefits for users whose target items are not covered by such memorized transitions.
We hypothesize that improving performance for these users requires LLMs to learn richer and predictive item--item relations complementing one-hop transitions. 
These relations can provide additional connections between the user’s input items and the target item.
Note that this motivates our training strategy (§~\ref{sec:method}).
We also clarify that one-hop memorization benefits certain users (i.e., one-hop memorization-benefiting-users) and therefore should not be viewed as inherently detrimental. 
Instead, our methodological goal is to improve performance for the remaining users who cannot benefit from one-hop memorization alone. Further discussion is in Appendix~\ref{subapp:clarifyingimplication}.

\begin{figure}[t]
    \centering
    \includegraphics[width=0.95\linewidth]{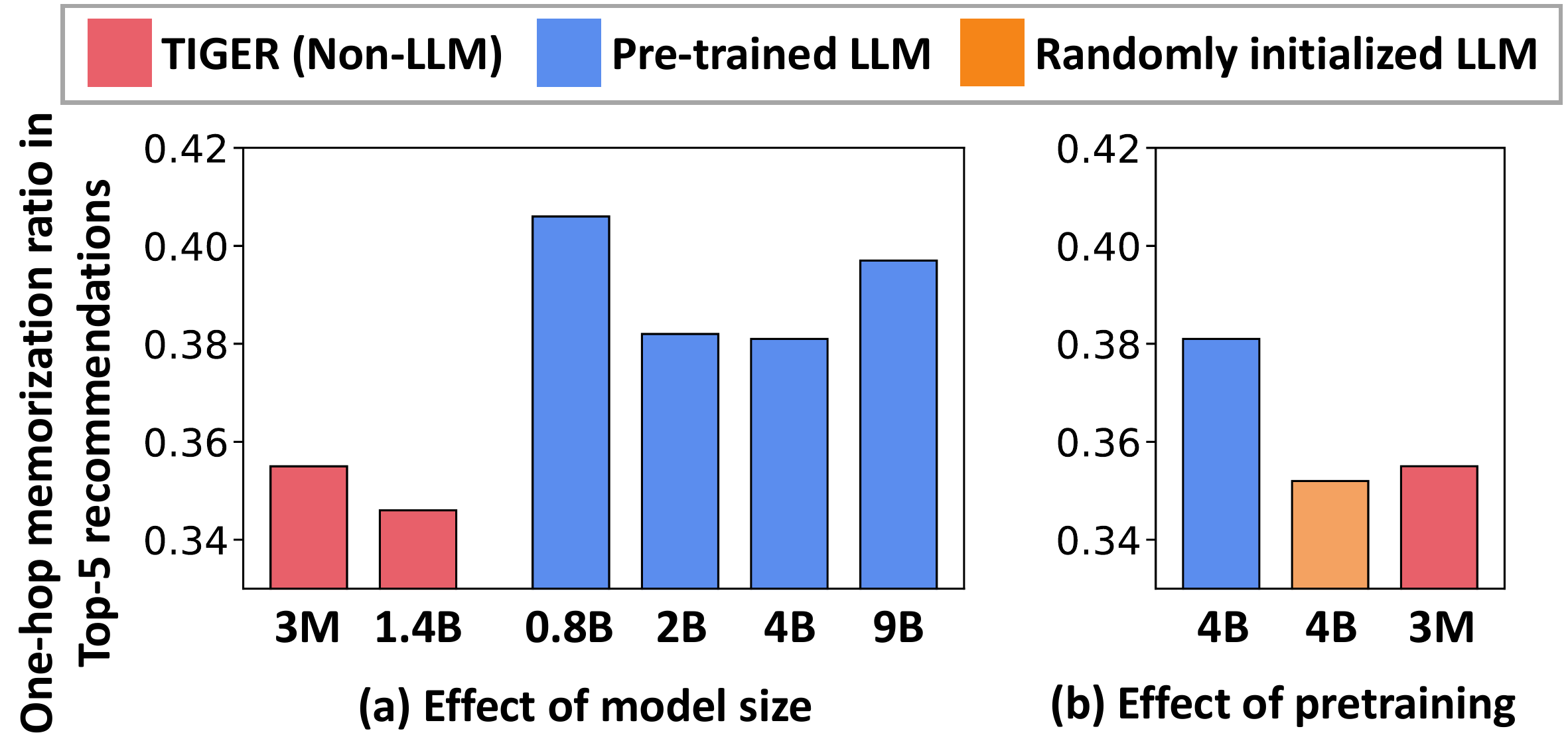}
    \caption{
    \textit{Potential factors behind one-hop memorization on the Sports dataset.}
    Model size shows no clear correlation with one-hop memorization (see (a)), whereas the pretrained LLM exhibits stronger reliance on one-hop memorization than its randomly initialized counterpart (see (b)).
    }
    \label{fig:underlyingfactor}    
\end{figure}

\subsection{Discussion of the underlying factors}
In this section, we discuss several potential factors underlying LLMs’ strong one-hop memorization behavior.
Specifically, we focus on three key factors that distinguish LLMs from non-LLM-based models: (1) model size, (2) pretraining, (3) model architectures.

\noindent{\textbf{Model size analysis.}
To analyze the impact of model size, we additionally consider models across a diverse range of scales: TIGER, a non-LLM-based model scaled up to 1.4B parameters, and Qwen-3.5 backbone LLMs of varying sizes.
As shown in Figure~\ref{fig:underlyingfactor}(a), model size exhibits no clear correlation with reliance on one-hop memorization. 
Specifically, increasing the size of either the non-LLM-based GR model TIGER or the backbone LLM does not consistently increase one-hop memorization. 
These results suggest that model size is unlikely to be the primary factor driving one-hop memorization.}

\noindent\textbf{Pretraining analysis.}
To examine whether pretraining is related to one-hop memorization, we additionally consider a randomly initialized LLM with the same architecture and size as its pretrained counterpart. 
As shown in Figure~\ref{fig:underlyingfactor}(b), the randomly initialized LLM exhibits substantially lower reliance on one-hop memorization than the pretrained model and shows a memorization ratio closer to that of the non-LLM-based GR model. 
This result reveals an association between pretrained initialization and stronger one-hop memorization, although it does not isolate pretraining as the causal factor.
However, the pretraining pipelines of many LLMs are not publicly disclosed, making it difficult to conduct further analysis or develop methods that directly intervene in the pretraining process. 
Thus, our method focuses on the post-training stage and can be applied to existing LLMs without modifying their pretraining procedures.
Notably, even the randomly initialized LLM generates valid item IDs at a rate comparable to its pretrained counterpart: 99\% of generated IDs map to existing items in both models.
Thus, its lower memorization ratio is unlikely to be explained by invalid outputs.

\noindent{\textbf{Architecture analysis.}
Note that the LLM results reported in \S~\ref{subsec:analysis-result} are based on models from the Qwen family~\cite{yang2025qwen3}.
To determine whether the observed behavior is specific to Qwen’s architecture, we extend our analysis to two additional LLM families that adopt distinct architectural choices, including different normalization schemes: LLaMA~\cite{grattafiori2024llama} and Phi~\cite{abouelenin2025phi}.
Our results show that reliance on one-hop memorization is consistently observed across these models, suggesting that this behavior is not limited to a particular LLM architecture; detailed results are provided in \S~\ref{subsec:llmgeneralization}.}

\begin{figure*}[!t]
    \centering
    \includegraphics[width=0.9\textwidth]{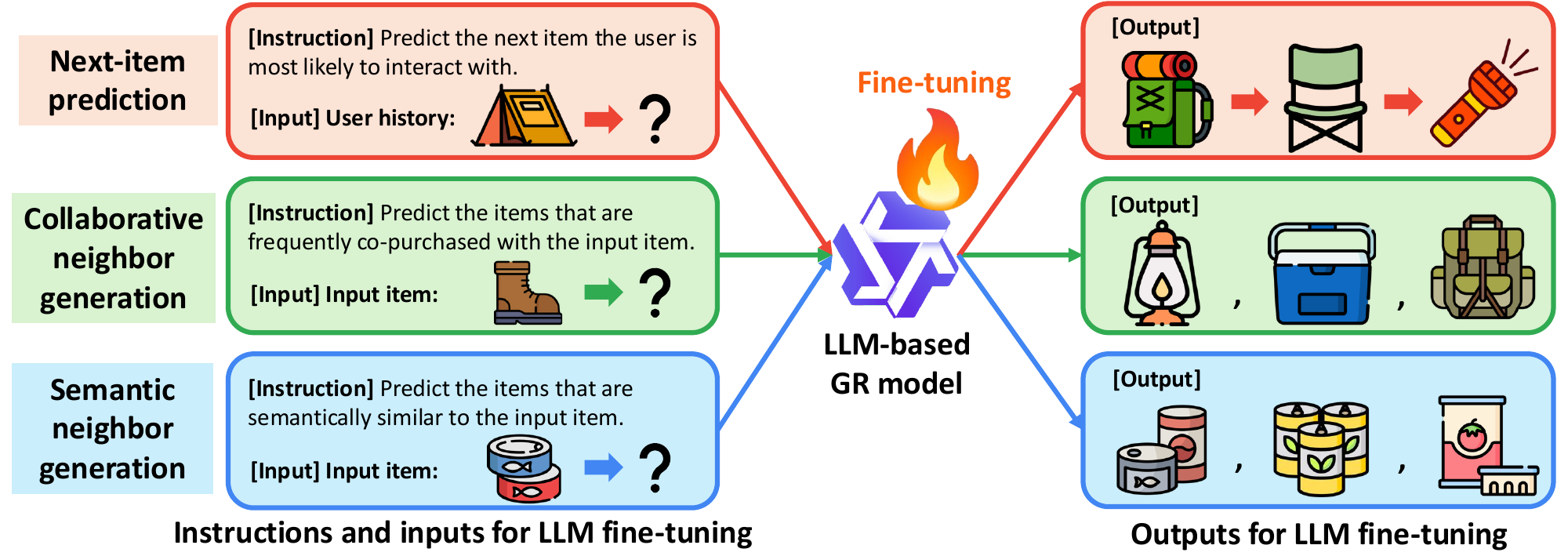}    
    \caption{
    \textit{Overview of \method.}
    To train LLM-based GR models, \method uses three tasks and jointly optimizes them: 
    (1) next-item prediction (\nextbg{red box}), 
    (2) collaborative neighbor generation for each item (\collabbg{green box}), 
    and (3) semantic neighbor generation for each item (\semanticbg{blue box}).
    }
    \label{fig:method}
\end{figure*}

    \section{Proposed method: \method}
    \label{sec:method}

Motivated by our analysis in \S~\ref{sec:analysis}, which implies that one-hop transitions alone are insufficient for certain users and need to be complemented by richer item--item relations, we introduce \method (\textbf{\underline{I}}tem–\textbf{\underline{I}}tem 
\textbf{\underline{R}}elation 
\textbf{\underline{G}}eneration), 
which jointly trains LLMs for next-item prediction and the explicit generation of \textit{multi-hop collaborative} and \textit{semantic} neighbors.
Figure~\ref{fig:method} provides a visual overview of \method.
We first motivate and summarize \method (§~\ref{subsec:method-intuition}), then detail its design (§~\ref{subsec:method-detail}).

\subsection{Overview and intuition for \method}\label{subsec:method-intuition}

\noindent\textbf{Overview.}
For each item, we construct two sets of ``neighbors'' to use as item-to-item supervision signals in our auxiliary training tasks.
Specifically, for an item $i_k$, we select (1) {{\em collaborative neighbors} from items that frequently co-occur with $i_k$ in a multi-hop window within user sequences}, and (2) {\em semantic neighbors} as items with similar textual descriptions.
Then, while training the LLM to predict next items in user sequences, we also train the LLM to generate the corresponding neighbor sets for each item $i_k\in\mathcal{I}$, enabling it to explicitly learn the relationships between $i_{k}$ and its neighbors.

\noindent\textbf{Key intuition.}
Our key criteria in selecting item-to-item supervision signals are whether such relations provide strong predictive information for the target items while remaining distinct from one-hop transitions, as motivated by the implications from our analysis (§~\ref{subsec:analysis-result}).
We provide our intuition on how our item-to-item supervision aligns with this goal.

First, for collaborative neighbors, frequently \textit{co-purchased} items provide strong evidence that interacting with one item increases the likelihood of interacting with the other~\cite{sarwar2001item}.
However, items far apart in a user sequence may reflect different intents and introduce noise. 
To reduce it, we define collaborative neighbors as items co-occurring within a multi-hop time window.

Second, for semantic neighbors, \textit{semantic relevance} captures item--item associations useful for recommendation, such as complementarity and/or theme-consistent relations. 
For instance, if a LEGO enthusiast shows interest in “\texttt{LEGO City Race Car}”, “\texttt{LEGO City Camper Van}” is a high-fidelity semantically grounded recommendation candidate due to the items' shared theme. 

\noindent\textbf{Evidence from data analysis.}
To verify whether our collaborative neighbors and semantic neighbors achieve our key objectives, we provide a data analysis in Appendix~\ref{subapp:data_analysis}. 
In short, we show that (1) these two types of neighbors provide strong predictive signals for users’ target items, comparable to or even stronger than one-hop transitions, and (2) their overlap with one-hop transitions is limited, suggesting that they provide additional, non-redundant signals complementing one-hop transitions.

\subsection{Design details of \method}
\label{subsec:method-detail}

\noindent\textbf{Obtaining collaborative neighbors.}
For collaborative neighbors, we first define an item--item pair weight as the frequency with which they co-appear within a specified time window in user sequences.
Specifically, the item--item pair weight $w_{kq}$ between items $i_k$ and $i_q$ is defined as follows:
\begin{align*}
    &w_{kq} = \sum_{u\in \mathcal{U}}\mathbb{I}
    [
    i_{k} \sim^{(u)}_{W} i_{q}
    ], \quad  \ i_{k} \sim^{(u)}_{W} i_{q} \iff \\
    &\exists t,j\in [\vert s_{u}\vert]:i^{(u)}_{t} = i_{k},i^{(u)}_{j}=i_{q},0<\vert t-j\vert \leq W,
\end{align*}
where $\mathbb{I}[\cdot]$ is the indicator function and $W$ denotes the time-window length.
Lastly, the collaborative neighbor sequence of item $i_{k}$, denoted by $n_{k}$, is defined as the sequence of top$-N_{1}$ items sorted in descending order of their pairwise weights $w_{kq}$ among all $i_{q} \in \mathcal{I}\setminus \{i_{k}\}$.
We also consider excluding one-hop transitions when obtaining collaborative neighbors, but this variant underperforms our approach; additional discussion is in Appendix~\ref{subapp:onehopout}.

\noindent\textbf{Obtaining semantic neighbors.}
For semantic neighbors, we first encode each item’s textual description into an embedding.
Specifically, for each item $i_{k} \in \mathcal{I}$, we use a pre-trained language embedding model $f$ to obtain its embedding $\mathbf{z}_{k} = f(\texttt{t}(i_{k}))$, where $\texttt{t}(i_{k})$ is a textual description of $i_{k}$.
Lastly, the semantic neighbor sequence of item $i_{k}$, denoted by $m_{k}$, is defined as the sequence of top$-N_{2}$ items sorted in descending order of their cosine similarities (i.e., $(\mathbf{z}^{T}_{k}\mathbf{z}_{q}) / (\|\mathbf{z_{k}}\| \|\mathbf{z_{q}}\|)$) for each item $i_{k} \in \mathcal{I}$ among all $i_{q} \in \mathcal{I} \setminus \{i_{k}\}$.
Alternatively, instead of performing embedding-based full search over all items, we can efficiently retrieve semantic neighbors using semantic-ID-based hashing, while still achieving competitive performance; the detailed experimental setup and results for this alternative design are provided in Appendix~\ref{subapp:efficient_semantic}.

\noindent\textbf{Training objective.}
After obtaining collaborative neighbors $n_{k}$ and semantic neighbors $m_{k}$ for each item $ i_{k}\in \mathcal{I}$, we train an LLM to capture these relations by learning to generate the corresponding neighbors for a given $i_{k}$.
To this end, as in next-item prediction (\S~\ref{subsec:analysis-setup}), each item $i_{k} \in \mathcal{I}$ is represented by its ID \texttt{id}$(i_{k})$ (either an SID or TID) and title \texttt{t}$(i_{k})$. 
We then construct input prompts for the collaborative-neighbor and semantic-neighbor generation tasks; examples of these prompts are in Appendices~\ref{subapp:col-neigh-prompt} and~\ref{subapp:sem_neigh-prompt}.

Lastly, we perform supervised fine-tuning of the LLM. 
Formally, let $\mathcal{T}^{(n_k)}$ and  $\mathcal{T}^{(m_k)}$ denote the tokenized representations of the sequences $n_k$ and $m_k$, respectively.
Similarly, let $\mathcal{T}^{(i_{k})}$ denote the tokenized representation of input item $i_{k}$.
Then, the LLM training loss for item $i_{k}$'s (1) collaborative neighbor generation $\mathcal{L}^{(C)}_{k}$ and (2) semantic neighbor generation $\mathcal{L}^{(S)}_{k}$ is defined as follows:

\begin{align}
    &\mathcal{L}^{(C)}_{k}=
\sum_{t=1}^{\left|\mathcal{T}^{(n_k)}\right|}
\frac{\log p_{\theta}
\left(
\mathcal{T}^{(n_k)}_{t}
\mid
\mathcal{T}^{(n_k)}_{< t}, \mathcal{T}^{(i_{k})},\mathcal{P}
\right)}{-{\vert \mathcal{T}^{(n_k)}\vert}},\label{eq:collaborative-loss} 
\\
    &\mathcal{L}^{(S)}_{k}=
\sum_{t=1}^{\left|\mathcal{T}^{(m_k)}\right|}
\frac{\log p_{\theta}
\left({
\mathcal{T}^{(m_k)}_{t}
\mid
\mathcal{T}^{(m_k)}_{< t}, \mathcal{T}^{(i_{k})}, \mathcal{S}
}\right)}{-{\vert \mathcal{T}^{(m_k)}\vert}},\label{eq:semantic-loss}
\end{align}
where $p_{\theta}$ denotes the token-generation probability of the LLM parameterized by $\theta$, $\mathcal{T}_{t}$ is the $t-$th token of the sequence $\mathcal{T}$ with $\mathcal{T}_{0}=\emptyset$, and $\mathcal{P}$ and $\mathcal{S}$ denote the token representations of the instructions for collaborative neighbor generation and semantic neighbor generation, respectively.

\noindent\textbf{Altogether, \method.}
Overall, \method jointly trains an LLM on three tasks: 
(1) next-item prediction, optimized with $\mathcal{L}_{u}$ (Eq.~\eqref{eq:nextitempredloss}); 
(2) collaborative-neighbor generation; 
and (3) semantic-neighbor generation.\footnote{One alternative is to use the neighbor generation tasks for pre-training, but this approach underperforms \method.
Further discussion is in Appendix~\ref{subapp:whyjoint}.}
Each training batch contains a mixture of samples from the three tasks.
Formally, let $\mathcal{U}_{(b)} \subseteq \mathcal{U}$ denote the users in batch $b$ for next-item prediction, and let $\mathcal{I}^{(C)}_{(b)} \subseteq \mathcal{I}$, and $\mathcal{I}^{(S)}_{(b)} \subseteq \mathcal{I}$ denote the items in batch $b$ for which collaborative neighbors and semantic neighbors are generated, respectively.
Then, for batch $b$, \method updates the parameters of the LLM via gradient descent by minimizing the following loss:
\begin{equation*}\label{eq:finalloss}
    \mathcal{L} = 
    \sum_{u\in\mathcal{U}_{(b)}} \frac{\mathcal{L}_{u}}{B}  + \sum_{i_{k}\in \mathcal{I}^{(C)}_{(b)}} \frac{\lambda_1\mathcal{L}^{(C)}_{k}}{B} + 
    \sum_{i_{t}\in \mathcal{I}^{(S)}_{(b)}} \frac{\lambda_2\mathcal{L}^{(S)}_{t}}{B},
\end{equation*}
where $B$ denotes the batch size, and $\lambda_1$ and $\lambda_2$ denote the hyperparameters that control the effects of the collaborative- and semantic-neighbor generation losses, respectively.

\begin{table*}[t!]
\centering
\setlength{\tabcolsep}{5pt}
\renewcommand{\arraystretch}{0.85} 
\caption{\textit{Main recommendation performance.}
R@K and N@K denote Recall@K and NDCG@K, respectively.
\bestbox{\textbf{Highlighted}} cells indicate the best-performing results.
Across all cases, \method achieves the best recommendation performance.
}
\label{tab:main_results}
\resizebox{0.9\textwidth}{!}{
\begin{tabular}{l | cccc | cccc | cccc}
\toprule
\multirow{2}{*}[-2pt]{\textbf{Methods}} & \multicolumn{4}{c|}{\textbf{Sports}} & \multicolumn{4}{c|}{\textbf{Toys}} & \multicolumn{4}{c}{\textbf{Beauty}} \\
\cmidrule(lr){2-5} \cmidrule(lr){6-9} \cmidrule(lr){10-13}
 & R@5 & N@5 & R@10 & N@10 & R@5 & N@5 & R@10 & N@10 & R@5 & N@5 & R@10 & N@10 \\
\midrule
\midrule
\multicolumn{13}{c}{\textit{Traditional recommender systems}} \\
\midrule
LightGCN &
0.0247 & 0.0162 & 0.0382 & 0.0205 & 
0.0371 & 0.0241 & 0.0556 & 0.0300 & 
0.0359 & 0.0236 & 0.0590 & 0.0311 \\

SimGCL &
0.0257 & 0.0168 & 0.0394 & 0.0212 & 
0.0387 & 0.0253 & 0.0581 & 0.0315 & 
0.0377 & 0.0239 & 0.0595 & 0.0309 \\

SASRec &
0.0217 & 0.0137 & 0.0356 & 0.0182 & 
0.0482 & 0.0314 & 0.0725 & 0.0392 & 
0.0373 & 0.0225 & 0.0624 & 0.0306 \\

FDSA &
0.0237 & 0.0156 & 0.0358 & 0.0196 & 
0.0564 & 0.0408 & 0.0764 & 0.0472 & 
0.0435 & 0.0289 & 0.0650 & 0.0358 \\

S$^3$-Rec &
0.0230 & 0.0151 & 0.0362 & 0.0194 & 
0.0551 & 0.0375 & 0.0777 & 0.0448 & 
0.0465 & 0.0307 & 0.0699 & 0.0383 \\

\midrule
\multicolumn{13}{c}{\textit{Non-LLM-based GR models}} \\
\midrule

TIGER &
0.0267 & 0.0170 & 0.0409 & 0.0216 & 
0.0484 & 0.0335 & 0.0727 & 0.0413 & 
0.0492 & 0.0327 & 0.0728 & 0.0402 \\

LETTER &
0.0260 & 0.0169 & 0.0413 & 0.0217 & 
0.0502 & 0.0331 & 0.0735 & 0.0408 & 
0.0498 & 0.0334 & 0.0732 & 0.0409 \\

\midrule
\multicolumn{13}{c}{\textit{Auxiliary-training-based GR models}} \\
\midrule

P5 &
0.0304 & 0.0201 & 0.0470 & 0.0249 & 
0.0545 & 0.0380 & 0.0778 & 0.0449 & 
0.0512 & 0.0350 & 0.0717 & 0.0415 \\

ReAT &
0.0295 & 0.0199 & 0.0425 & 0.0239 & 
0.0587 & 0.0409 & 0.0833 & 0.0485 & 
0.0559 & 0.0386 & 0.0775 & 0.0456 \\

LC-Rec &
0.0317 & 0.0206 & 0.0485 & 0.0258 & 
0.0615 & 0.0422 & 0.0879 & 0.0506 & 
0.0620 & 0.0454 & 0.0843 & 0.0523 \\

EAGER-LLM &
0.0373 & 0.0251 & 0.0569 & 0.0315 & 
0.0546 & 0.0359 & 0.0788 & 0.0437 & 
0.0548 & 0.0369 & 0.0830 & 0.0459 \\

\midrule
\multicolumn{13}{c}{\textit{LLM-based GR models}} \\
\midrule

PLUM &
0.0297 & 0.0199 & 0.0436 & 0.0245 & 
0.0549 & 0.0389 & 0.0737 & 0.0449 & 
0.0522 & 0.0360 & 0.0745 & 0.0433 \\

OneRec-Think &
0.0288 & 0.0199 & 0.0412 & 0.0239 & 
0.0579 & 0.0412 & 0.0797 & 0.0482 & 
0.0563 & 0.0398 & 0.0791 & 0.0471 \\

GRAM &
0.0375 & 0.0256 & 0.0554 & 0.0314 & 
0.0718 & 0.0516 & 0.0987 & 0.0603 & 
0.0641 & 0.0451 & 0.0890 & 0.0531 \\

GRLM &
0.0375 & 0.0260 & 0.0539 & 0.0313 & 
0.0684 & 0.0477 & 0.0942 & 0.0561 & 
0.0607 & 0.0430 & 0.0846 & 0.0506 \\

Naive-TID & 
0.0337 & 0.0226 & 0.0482 & 0.0273 & 
0.0634 & 0.0438 & 0.0892 & 0.0521 & 
0.0606 & 0.0429 & 0.0828 & 0.0501 \\

AgenticTagger &
0.0299 & 0.0194 & 0.0474 & 0.0251 & 
0.0540 & 0.0380 & 0.0781 & 0.0457 & 
0.0492 & 0.0332 & 0.0728 & 0.0408 \\

\midrule

\method (Ours) &
\best{0.0406} & \best{0.0283} & \best{0.0585} & \best{0.0341} & 
\best{0.0780} & \best{0.0541} & \best{0.1097} & \best{0.0642} & 
\best{0.0726} & \best{0.0513} & \best{0.1011} & \best{0.0606} \\
\bottomrule
\end{tabular}
}
\vspace{-2mm}
\end{table*}

    
    \section{Experiments}
    \label{sec:experiment}

In this section, we evaluate the effectiveness of \method in the recommendation tasks.
To this end, we investigate the following five research questions:
\begin{itemize}

\item \textbf{\textit{(RQ1) General recommendation performance.}} How effective is \method for the recommendation task?

\item \textbf{\textit{(RQ2) Generalization across item IDs.}} Does the effectiveness of \method generalize across different item ID assignments?

\item \textbf{\textit{(RQ3) Generalization across LLM families.}} Do our key observations and the effectiveness of \method generalize across diverse LLM families?

\item \textbf{\textit{(RQ4) Achievement of design goal.}} Does \method achieve its intended design objective?

\item \textbf{\textit{(RQ5) Ablation study.}} Do all the key components of \method contribute to its strong performance?

\end{itemize}

\subsection{Experimental setting}\label{subsec:experimentsetting}

\noindent\textbf{Datasets.}
For experiments, we use three benchmark sequential recommendation domains from the Amazon Reviews dataset~\cite{he2016ups}: Sports, Toys, and Beauty. 
Additional results on other domain datasets are in Appendix~\ref{subapp:yelpresult}.

\noindent\textbf{Evaluation.}
Following common practice in sequential recommendation~\cite{kang2018self, geng2022recommendation}, we use the \textit{leave-one-out} protocol. 
Specifically, for each user sequence, we use the last item for testing, the second-to-last item for validation, and the remaining interactions for training.
We consider the \textit{full-item ranking setting}, where all items are considered as possible recommendation candidates.
For metrics, we use Recall@$K$ and NDCG@$K$, where $K\in\{5,10\}$.

\noindent\textbf{Baseline methods.}
We compare \method against 17 baselines from four categories.
Specifically, we use
(1) five traditional models 
(LightGCN~\cite{he2020lightgcn}, 
SimGCL~\cite{yu2022graph},
SASRec~\cite{kang2018self}, 
FDSA~\cite{zhang2019feature}, 
S$^{3}$-Rec~\cite{zhou2020s3}), 
(2) two non-LLM-based GR models 
(TIGER~\cite{rajput2023recommender}, LETTER~\cite{wang2024learnable}), 
(3) four GR methods that use joint training of auxiliary tasks and next-item prediction (P5~\cite{geng2022recommendation}, 
ReAT~\cite{cao2024aligning}, 
LC-Rec~\cite{zheng2024adapting}, 
EAGER-LLM~\cite{hong2025eager}), and
(4) six LLM-based GR models using SIDs  
(PLUM~\cite{he2026plum}, 
OneRec-Think~\cite{liu2025onerec}) or TIDs
(Naive (LLM trained solely with next-item prediction), 
GRAM~\cite{lee2025gram},
GRLM~\cite{zhang2026unleashing}, AgenticTagger~\cite{xie2026agentictagger}). 
Further baseline details are in Appendix~\ref{subapp:baselines}.

\noindent\textbf{Proposed method: \method.}
We fully fine-tune the backbone LLM as described in \S~\ref{subsec:method-detail}. 
We use TIDs as the default item IDs, except when evaluating the effectiveness of \method under SIDs in \S~\ref{subsec:sidgeneralization}. 
Note that each user’s target item is explicitly excluded from the user sequence when obtaining collaborative neighbors to prevent data leakage.
Item ID details and hyperparameter configurations of \method are in Appendices~\ref{subapp:ourtermid} and~\ref{subapp:methoddetail}, respectively.

\subsection{Recommendation performance}\label{subsec:mainperformance}
\noindent\textbf{Setup.}
For non-LLM baselines, we use the backbone models from their original papers. 
For LLM-based baselines, we use Qwen-3.5-4B~\cite{yang2025qwen3}, the same backbone as \method, and fully fine-tune it. 
We use the item identifiers proposed in each paper, except for LLM-based GR models with auxiliary training, for which we use the same identifiers as \method for a fair comparison of training strategies. 
When our reproduced results are lower than the originally reported results, we use the results reported in the corresponding paper, as their methods were also evaluated under the same setting.
In addition, we further analyze cold-start item recommendation in Appendix~\ref{subapp:coldstart}.

\begin{figure*}[t]
    \centering
    \includegraphics[width=\linewidth]{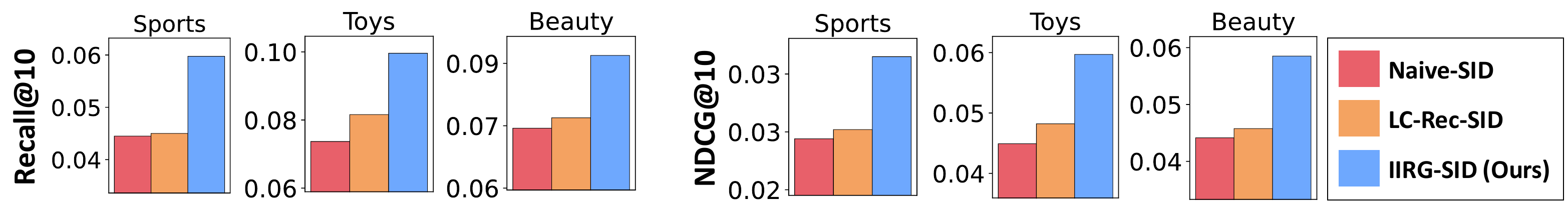}
    \caption{\textit{Generalization to SIDs.}
    \method remains effective under SIDs (\method-SID), improving over the LLM trained only with next-item prediction (Naive-SID) and outperforming the strongest SID-based baseline (LC-Rec-SID).
    }
    \label{fig:main_sid}
    \vspace{2mm}
\end{figure*}

\noindent\textbf{Result.}
As shown in Table~\ref{tab:main_results}, \method consistently performs best across all datasets and metrics. 
Two points stand out.
First, \method{} consistently improves over the LLM trained solely with next-item prediction (Naive), achieving a 21\% average improvement in Recall@5 across datasets.
Second, \method outperforms all baseline training methods across all datasets, demonstrating that it is a more effective training method than existing LLM-based GR training baselines.

\subsection{Generalization across item IDs}\label{subsec:sidgeneralization}

\noindent\textbf{Setup.} 
We showed the effectiveness of \method under TIDs in \S~\ref{subsec:mainperformance}. 
To examine whether this improvement generalizes to SIDs, we repeat the experiments from \S~\ref{subsec:mainperformance} using SIDs as item IDs. 
For a fair comparison, Naive, LC-Rec, and \method all use the SIDs from~\citet{he2026plum}; we use LC-Rec as the strongest baseline among those to which SIDs are applicable.\footnote{Naive with SIDs corresponds to PLUM~\cite{he2026plum}.}
Note that LC-Rec uses more training tokens than \method.
Additional results with different SIDs are provided in Appendix~\ref{subapp:rkmeans}.

\noindent\textbf{Result.}
As shown in Figure~\ref{fig:main_sid}, \method consistently improves over Naive and outperforms LC-Rec, the strongest applicable SID baseline, across all datasets. 
This result indicates that the effectiveness of \method is not limited to term IDs but also extends to semantic IDs.

\begin{figure}[t]
    \centering
    \includegraphics[width=\linewidth]{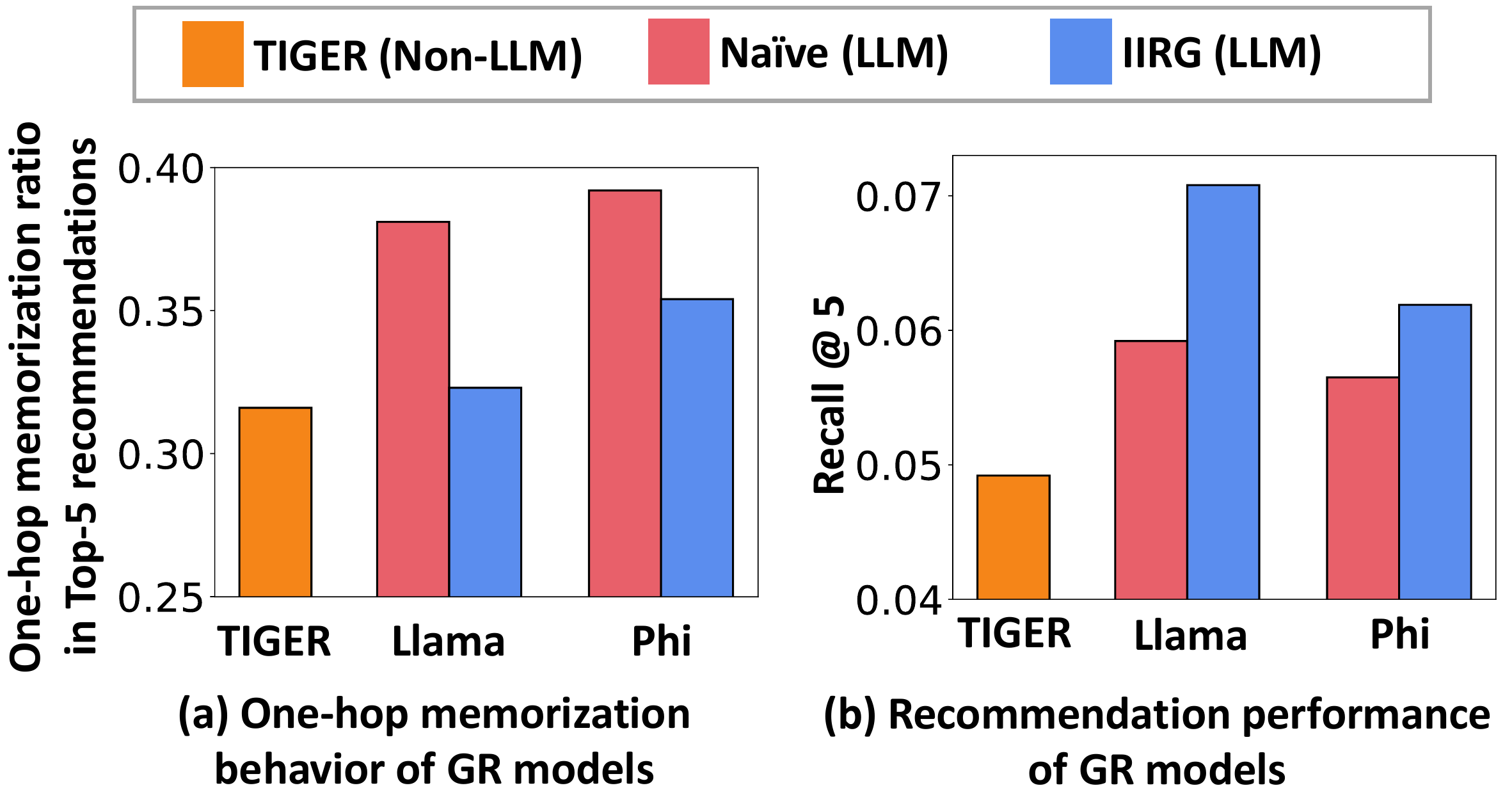}
    \caption{
    \textit{Generalization across LLMs on the Beauty dataset.}
    As shown in (a), two backbone models from distinct LLM families, Llama-3.2-3B and Phi-4-3.8B, exhibit stronger one-hop memorization than the non-LLM-based TIGER model when trained solely with next-item prediction (Naive). 
    Notably, \method consistently (1) reduces reliance on one-hop memorization (see (a)), and (2) improves recommendation performance (see (b)), across both LLM families.
    }
    \label{fig:llmgeneralization}
\end{figure}

\begin{figure}[t]
    \centering
    \includegraphics[width=\linewidth]{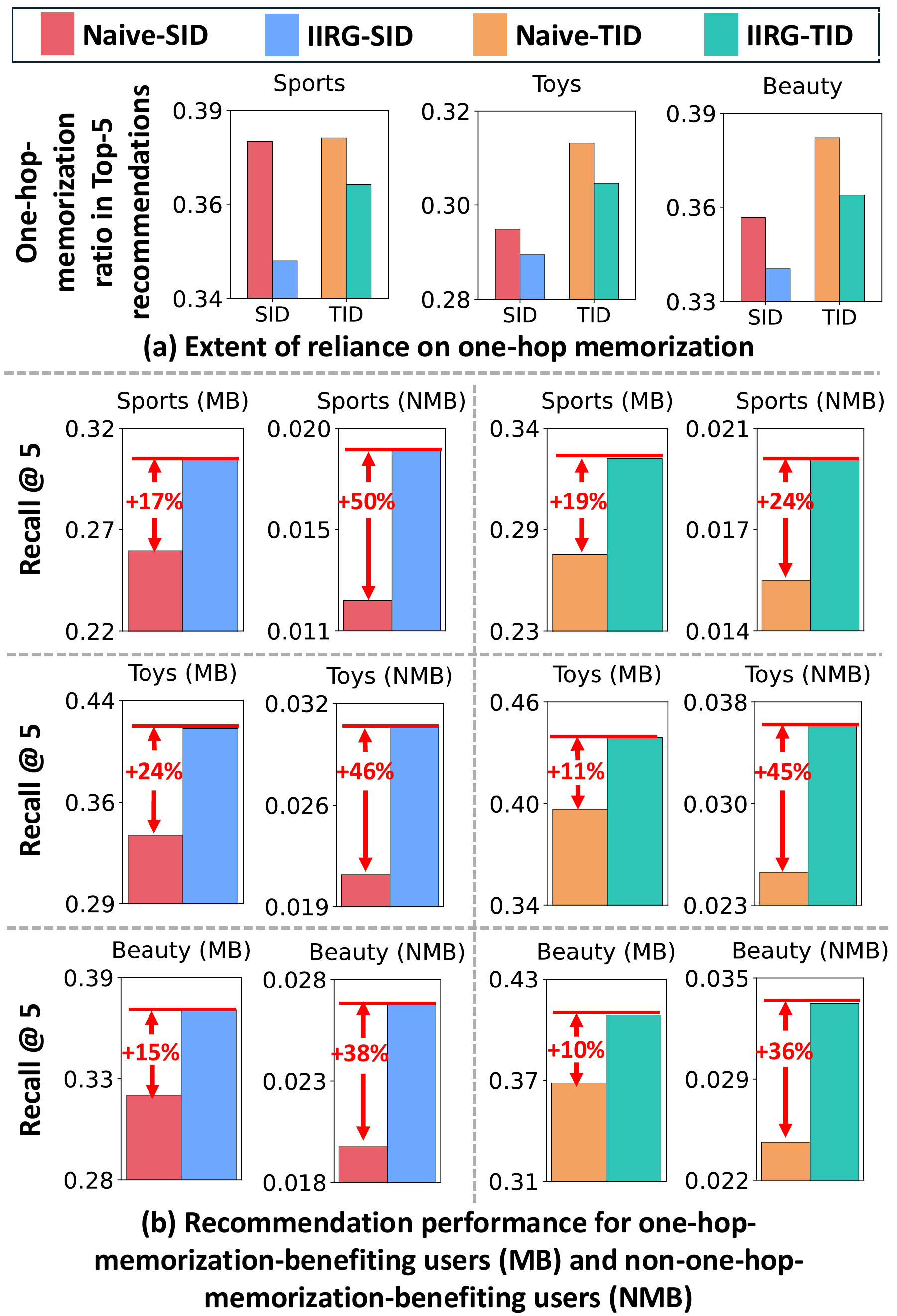}
    \caption{
    \textit{Achievement of design goal.}
    \method{} reduces Naive's reliance on one-hop memorization, an LLM trained solely with next-item prediction (a), and yields larger percentage gains for non-one-hop-memorization-benefiting users than for one-hop-memorization-benefiting users (b). 
    These trends hold across datasets and ID types.
    }
    \label{fig:design_objective}
\end{figure}

\subsection{Generalization across LLM families}\label{subsec:llmgeneralization}

\noindent\textbf{Setup.} 
{Recall that our analysis of one-hop memorization (\S~\ref{sec:analysis}) and recommendation performance (\S~\ref{subsec:mainperformance} and \S~\ref{subsec:sidgeneralization}) primarily relied on Qwen-based LLMs. 
In this subsection, we examine whether the observed one-hop memorization behavior and the effectiveness of \method generalize across diverse LLM families.
To this end, we employ two additional LLMs, namely Llama-3.2-3B~\cite{grattafiori2024llama} and Phi-4-3.8B~\cite{abouelenin2025phi}, as backbone models for generative recommendation.
Using TIDs as item IDs, we evaluate each model in terms of (1) one-hop memorization, following the setup in \S~\ref{subsec:analysis-setup}, and (2) recommendation performance when trained with either next-item prediction alone (Naive) or \method, following the setup in \S~\ref{subsec:mainperformance}.
We use the Beauty dataset for investigation.}

\begin{table*}[t]
\centering
\renewcommand{\arraystretch}{0.9}
\caption{\textit{Ablation study.}
{Collab.} and {Semantic} indicate the use of collaborative neighbor generation and semantic neighbor generation, respectively.
\method, which uses both generation tasks, achieves the strongest performance.
}
\label{tab:ablation_results}
\resizebox{\textwidth}{!}{
\begin{tabular}{cc | cccc | cccc | cccc}
\toprule
\multirow{2}{*}[-2pt]{\textbf{Collab.}} 
& \multirow{2}{*}[-2pt]{\textbf{Semantic}} 
& \multicolumn{4}{c|}{\textbf{Sports}} 
& \multicolumn{4}{c|}{\textbf{Toys}} 
& \multicolumn{4}{c}{\textbf{Beauty}} \\
\cmidrule(lr){3-6} \cmidrule(lr){7-10} \cmidrule(lr){11-14}
& & R@5 & N@5 & R@10 & N@10 
& R@5 & N@5 & R@10 & N@10 
& R@5 & N@5 & R@10 & N@10 \\
\midrule

\xmark & \xmark &
0.0337 & 0.0226 & 0.0482 & 0.0273 & 
0.0634 & 0.0438 & 0.0892 & 0.0521 & 
0.0606 & 0.0429 & 0.0828 & 0.0501 \\

\cmark & \xmark &
0.0401 & 0.0274 & 0.0568 & 0.0328 &
0.0703 & 0.0490 & 0.0975 & 0.0577 & 
0.0671 & 0.0475 & 0.0929 & 0.0558 \\

\xmark & \cmark &
0.0391 & 0.0263 & 0.0559 & 0.0323 & 
0.0745 & 0.0507 & 0.1038 & 0.0604 & 
0.0663 & 0.0464 & 0.0941 & 0.0554 \\

\midrule

\cmark & \cmark &
 \textbf{0.0406} &   \textbf{0.0283} &   \textbf{0.0585} &   \textbf{0.0341} & 
  \textbf{0.0780} &   \textbf{0.0541} &   \textbf{0.1097} &   \textbf{0.0642} & 
  \textbf{0.0726} &   \textbf{0.0513} &   \textbf{0.1011} &   \textbf{0.0606} \\

\bottomrule
\end{tabular}
}
\vspace{2mm}
\end{table*}

\noindent\textbf{Result.}
{The analysis results are presented in Figure~\ref{fig:llmgeneralization}.
First, both additional LLM families exhibit stronger one-hop memorization than TIGER, a non-LLM-based model. 
This result is consistent with our findings in \S~\ref{subsec:analysis-result} (see Figure~\ref{fig:llmgeneralization}(a)), indicating that the observations reported in \S~\ref{sec:analysis} are not specific to the Qwen family but generalize to other LLM families.
Second, \method consistently (1) reduces Naive's reliance on one-hop memorization (see Figure~\ref{fig:llmgeneralization}(a)) and (2) improves recommendation performance over Naive LLMs (see Figure~\ref{fig:llmgeneralization}(b)). 
These results suggest that the effectiveness of \method generalizes across diverse LLM families.}

\subsection{Design goal achievement}\label{subsec:designgoal}
\noindent\textbf{Setup.}
We examine whether \method{} reduces the reliance on one-hop memorization exhibited by an LLM trained solely with next-item prediction, and whether it achieves stronger performance gains for users who do not benefit from one-hop memorization.
To this end, we repeat the analysis in §~\ref{sec:analysis} for \method{} and compare its results against Naive, an LLM trained solely with next-item prediction.

\noindent\textbf{Result.}
As shown in Figure~\ref{fig:design_objective}(a), \method reduces the Naive LLM's reliance on one-hop memorization. 
Also, as shown in Figure~\ref{fig:design_objective}(b), \method{} yields an average gain over Naive of 40\% for users who do not benefit from one-hop memorization, versus 16\% for users who do, across datasets and item IDs.
This suggests that \method provides strong gains for non-memorization-benefiting users, aligning with our key objective.

\subsection{Ablation study}

\noindent\textbf{Setup.}
We analyze how each neighbor-generation task contributes to \method's performance.
We remove collaborative neighbor generation, semantic neighbor generation, or both.
TIDs are used as item IDs.
In Appendix~\ref{subapp:additionalablation}, we further study (1) a variant trained on random neighbors and (2) heuristics that directly use collaborative or semantic neighbors as recommendation results, both of which underperform \method.

\noindent\textbf{Result.}
As shown in Table~\ref{tab:ablation_results}, \method consistently outperforms its variants across the datasets. 
These results demonstrate that the two neighbor-generation tasks are complementary and that combining them yields the strongest overall performance.

    \section{Conclusion}
    \label{sec:conclusion}

In this work, we investigate one-hop memorization behavior in LLM-based GR. 
We first show that LLMs rely more heavily on one-hop memorization than non-LLM-based GR models, and that their performance gains over these models are largely concentrated among users who benefit from such memorization. 
Motivated by this analysis, we propose \method{}, which jointly trains LLMs on next-item prediction and explicit generation of multi-hop collaborative and semantic neighbors.
Our experiments show that our item-to-item supervision consistently improves recommendation performance while yielding particularly large gains for users outside one-hop coverage.
    


    \bibliographystyle{ACM-Reference-Format}
	\bibliography{000Ref}

    \appendix

    \section{Details of item identifiers}
    \label{app:itemid}

Item identifiers are structured discrete representations of items that enable LLMs to memorize item-specific information and generate item recommendations.
Largely, item identifiers are categorized into: (1) semantic identifiers (semantic IDs) and (2) term-based identifiers (term IDs).

\subsection{Semantic identifiers}
Semantic identifiers (semantic IDs) are discrete code sequences that represent individual items.
For example, an item $i_{k}$ can be represented by a semantic ID sequence $(6,102,47)$.
Such semantic IDs are often generated from item metadata using discrete representation learning techniques, such as RQ-VAE~\cite{lee2022autoregressive} and Residual K-Means~\cite{zhou2025openonerec}.
Despite each semantic ID having an encoded meaning that captures item metadata information, LLMs do not have prior knowledge of these identifiers.
Therefore, instead of directly representing each item with integer sequences, methods typically do vocabulary expansion with special tokens for each unique integer at each position to represent the corresponding code, following prior studies~\cite{zheng2024adapting, he2026plum}.
Often, these special tokens are aligned with the LLM’s token space through a dedicated training procedure, such as generating the codes from a given item title~\cite{he2026plum}.

\subsection{Term-based identifiers}
One key challenge in using codebook-based identifiers (e.g., semantic IDs) for LLM-based generative retrieval is aligning the special tokens corresponding to the codes with the LLM’s token space, so that the LLM can effectively understand and use these codes~\cite{zhang2026unleashing}.
To avoid this cost, a promising research direction is to use terms (i.e., keywords that represent the characteristics of each item) to represent each item~\cite{tan2024idgenrec}.
For example, an item with title "\texttt{Retrospec Bicycles Urban/Fixed Gear Saddle}" can be represented as ["\texttt{Cycling}", "\texttt{Bicycle-part}", "\texttt{Saddle}", "\texttt{Fixed-gear}"].
In this way, existing language tokens can be naturally used to represent items, allowing LLMs to interpret them without additional alignment steps.

    \section{Prompt examples}
    \label{app:prompt}
    

\begin{figure*}[t]
\centering
\begin{promptblock}
\small
\textbf{Instruction:} 
Given a user's historical item interaction sequence, predict the keywords of the next item the user is most likely to interact with. 
Each item in the sequence is represented by exactly 4 keywords enclosed in square brackets []. 
The items are listed in chronological order.

\vspace{0.5em}
\textbf{Input:} Item keywords: [cycling, tire, wire-bead, gum-wall] Title: Cheng Shin C637 Road Bike Tire (Wire Bead, 27\&quot; x 1-1/4\&quot;, Black Wall).

\vspace{0.5em}
\textbf{Output:} 
Item keywords: [cycling, tube, 700mm, 35-40] Title: Sunlite Bicycle Tube, 700 x 35-40 (27 x 1-3/8) SCHRADER Valve.
\texttt{\textbackslash n}
Item keywords: [cycling, bicycle-parts, saddles, fixed-gear] Title: Retrospec Bicycles Urban/Fixed Gear Saddle.

\end{promptblock}
\caption{\textbf{Training sample of next-item prediction with term IDs.} 
Given the instruction and input, the LLM is trained to generate the output autoregressively using teacher forcing.}
\label{fig:termnextitemprompt}
\end{figure*}


\begin{figure*}[t]
\centering
\begin{promptblock}
\small
\textbf{Instruction:} 
Given a user's historical item interaction sequence, predict the next item the user is most likely to interact with. 
Each item in the sequence is represented by a unique identifier composed of 3 special tokens enclosed in square brackets []. 
The items are listed in chronological order.

\vspace{0.5em}
\textbf{Input:} Item keywords: $[<a_{195}><b_{311}><c_{230}>]$ Title: Cheng Shin C637 Road Bike Tire (Wire Bead, 27\&quot; x 1-1/4\&quot;, Black Wall).

\vspace{0.5em}
\textbf{Output:} Item keywords: $[<a_{84}><b_{500}><c_{133}>]$ Title: Sunlite Bicycle Tube, 700 x 35-40 (27 x 1-3/8) SCHRADER Valve.
\texttt{\textbackslash n}
Item keywords: $[<a_{546}><b_{11}><c_{132}>]$ Title: Retrospec Bicycles Urban/Fixed Gear Saddle.

\end{promptblock}
\caption{\textbf{Training sample of next-item prediction with semantic IDs.} 
Given the instruction and input, the LLM is trained to generate the output autoregressively using teacher forcing.}
\label{fig:semanticnextitemprompt}
\end{figure*}


\begin{figure*}[t]
\centering
\begin{promptblock}
\small
\textbf{Instruction:} 
Given a target item in the format [keywords, title], recommend five items that are most likely to be co-purchased with it.
\texttt{\textbackslash n} 
Return the items sorted by likelihood, from most likely to least likely, and format each item as [keywords, title].

\vspace{0.5em}
\textbf{Input:} 
Item keywords: [game, card-game, solitaire, mega-screen] Title: Mega Screen Solitaire.

\vspace{0.5em}
\textbf{Output:} 
Item keywords: [educational, stacking-activity, blocks, ten] Title: Fisher-Price Brilliant Basics Baby's First Blocks.
\texttt{\textbackslash n} 
Item keywords: [playsets, sports-equipment, basketball-sets, hoop-system] Title: Little Tikes EasyScore Basketball Set.
\texttt{\textbackslash n} 
Item keywords: [electronic, interactive, sound-based, farmer-theme] Title: Fisher-Price See 'n Say The Farmer Says.
\texttt{\textbackslash n} 
Item keywords: [game, board-games, card-games, busy-town] Title: Richard Scarry Busy Town.
\texttt{\textbackslash n}
Item keywords: [electronic, interactive, activity-center, musical-table] Title: LeapFrog Animal Adventure Learning Table (Frustration Free Packaging).

\end{promptblock}
\caption{\textbf{Training sample of collaborative neighbor generation with term IDs.} 
Given the instruction and input, the LLM is trained to generate the output autoregressively using teacher forcing.}
\label{fig:termcollaborativeprompt}
\end{figure*}


\begin{figure*}[t]
\centering
\begin{promptblock}
\small
\textbf{Instruction:} 
Given a target item in the format [keywords, title], recommend five items that are most likely to be co-purchased with it.
\texttt{\textbackslash n} 
Return the items sorted by likelihood, from most likely to least likely, and format each item as [keywords, title].

\vspace{0.5em}
\textbf{Input:} Item keywords: [$<a_{716}><b_{147}><c_{167}>$] Title: Mega Screen Solitaire.

\vspace{0.5em}
\textbf{Output:} Item keywords: [$<a_{431}><b_{440}><c_{64}>$] Title: Fisher-Price Brilliant Basics Baby's First Blocks. \texttt{\textbackslash n}
Item keywords: [$<a_{852}><b_{490}><c_{1}>$] Title: Little Tikes EasyScore Basketball Set. \texttt{\textbackslash n}
Item keywords: [$<a_{220}><b_{439}><c_{14}>$] Title: Fisher-Price See 'n Say The Farmer Says. \texttt{\textbackslash n}
Item keywords: [$<a_{160}><b_{98}><c_{214}>$] Title: Richard Scarry Busy Town. \texttt{\textbackslash n}
Item keywords: [$<a_{635}><b_{282}><c_{146}>$] Title: LeapFrog Animal Adventure Learning Table (Frustration Free Packaging).

\end{promptblock}
\caption{\textbf{Training sample of collaborative neighbor generation with semantic IDs.} 
Given the instruction and input, the LLM is trained to generate the output autoregressively using teacher forcing.}
\label{fig:semanticcollaborativeprompt}
\end{figure*}


\begin{figure*}[t]
\centering
\begin{promptblock}
\small
\textbf{Instruction:} 
Given a target item in the format [keywords, title], list five items that are most semantically similar to it.
\texttt{\textbackslash n}
Return the items sorted by similarity, from most similar to least similar, and format each item as [keywords, title].

\vspace{0.5em}
\textbf{Input:} Item keywords: 
[skin-care, cleansers, cream-cleansers, soap-free-formulation] Title: Burt's Bees Sensitive Facial Cleanser, 6 Fluid Ounces.

\vspace{0.5em}
\textbf{Output:} 
Item keywords: [skin-care, sensitive-skin, face-creams, night-creams] Title: Burt's Bees Sensitive Daily Moisturizing Cream, 1.8 Ounces.
\texttt{\textbackslash n}
Item keywords: [skin-care, facial-care, cleansers, orange-essence] Title: Burt's Bees Radiance Facial Cleanser, 6 Fluid Ounce.
\texttt{\textbackslash n}
Item keywords: [cleanser, face, cream, soap-bark] Title: Burt's Bees Soap Bark \&amp; Chamomile Deep Cleansing Cream 6 oz (Pack of 2).
\texttt{\textbackslash n}
Item keywords: [cleanser, face, soap-free, ultra-calming] Title: Aveeno Ultra-Calming Moisturizing Cream Cleanser, 5-Ounce Tubes  (Pack of 3).
\texttt{\textbackslash n}
Item keywords: [skin-care, facial-care, moisturizing, night-cream] Title: Burt's Bees Sensitive Night Cream, 1.8 Ounces.

\end{promptblock}
\caption{\textbf{Training sample of semantic neighbor generation with term IDs.} 
Given the instruction and input, the LLM is trained to generate the output autoregressively using teacher forcing.}
\label{fig:termsemanticprompt}
\end{figure*}


\begin{figure*}[t]
\centering
\begin{promptblock}
\small
\textbf{Instruction:} 
Given a target item in the format [keywords, title], list five items that are most semantically similar to it.
\texttt{\textbackslash n}
Return the items sorted by similarity, from most similar to least similar, and format each item as [keywords, title].

\vspace{0.5em}
\textbf{Input:} Item keywords: $[<a_{289}><b_{375}><c_{121}>]$ Title: Burt's Bees Sensitive Facial Cleanser, 6 Fluid Ounces.

\vspace{0.5em}
\textbf{Output:} 
Item keywords: $[<a_{269}><b_{406}><c_{35}>]$ Title: Burt's Bees Sensitive Daily Moisturizing Cream, 1.8 Ounces.
\texttt{\textbackslash n}
Item keywords: $[<a_{779}><b_{379}><c_{33}>]$ Title: Burt's Bees Radiance Facial Cleanser, 6 Fluid Ounce.
\texttt{\textbackslash n}
Item keywords: $[<a_{{779}}><b_{406}><c_{23}>]$ Title: Burt's Bees Soap Bark \&amp; Chamomile Deep Cleansing Cream 6 oz (Pack of 2).
\texttt{\textbackslash n}
Item keywords: $[<a_{609}><b_{1}><c_{2}>]$ Title: Aveeno Ultra-Calming Moisturizing Cream Cleanser, 5-Ounce Tubes  (Pack of 3).
\texttt{\textbackslash n}
Item keywords: $[<a_{269}><b_{302}><c_{93}>]$ Title: Burt's Bees Sensitive Night Cream, 1.8 Ounces.

\end{promptblock}
\caption{\textbf{Training sample of semantic neighbor generation with semantic IDs.} 
Given the instruction and input, the LLM is trained to generate the output autoregressively using teacher forcing.}
\label{fig:semanticsemanticprompt}
\end{figure*}

In this section, we present several examples for the prompts used in \method.
In the implementation, each sample is formatted using the Alpaca format adopted by LLaMA-Factory~\cite{zheng2024llamafactory}.

\subsection{Next-item prediction}\label{subapp:next-item-prompt}

The example data samples using term-based item identifiers and semantic identifiers are shown in Figure~\ref{fig:termnextitemprompt} and Figure~\ref{fig:semanticnextitemprompt}, respectively.
Given the instruction and anchor item (i.e., the first item the user has interacted with), the LLM is trained to generate the remaining parts of the user's interaction history.
This format is inspired by~\citet{zhang2026unleashing}.

\subsection{Collaborative neighbor generation}\label{subapp:col-neigh-prompt}

The example data samples using term-based item identifiers and semantic identifiers are shown in Figure~\ref{fig:termcollaborativeprompt} and Figure~\ref{fig:semanticcollaborativeprompt}, respectively.
Given the instruction and target item, the LLM is trained to generate the collaborative neighbors of the target item.

\subsection{Semantic neighbor generation}\label{subapp:sem_neigh-prompt}

The example data samples using term-based item identifiers and semantic identifiers are shown in Figure~\ref{fig:termsemanticprompt} and Figure~\ref{fig:semanticsemanticprompt}, respectively.
Given the instruction and target item, the LLM is trained to generate the semantic neighbors of the target item.

    \section{Experimental details}
    \label{app:expdetail}

In this section, we present further experimental details of our work.

\begin{table}[t]
\centering
\caption{\textbf{Dataset statistics.} Interactions denote the user-item interactions.}
\label{tab:dataset_statistics}
\begin{tabular}{l | rrrr}
\toprule
Dataset & \# Users & \# Items & \# Interactions\\
\midrule
Sports & 35,598 & 18,357 & 296,337 \\
Toys   & 19,412 & 11,924 & 167,597 \\
Beauty & 22,363 & 12,101 & 198,502 \\
\bottomrule
\end{tabular}
\end{table}



\subsection{Machines}\label{subapp:implementation}
Our work is conducted on Google Cloud Platform.
Each training job is run on a single GCP \texttt{a2-highgpu-8g} VM node equipped with 96 vCPUs, 680 GB of memory, and 8 NVIDIA A100 GPUs, each with 40 GB of GPU memory.
However, for training and evaluation of Qwen-3.5-9B backbone LLM, we use a single GCP \texttt{a2-ultragpu-8g} VM node equipped with 96 vCPUs, 1,360 GB of memory, and 8 NVIDIA A100 GPUs, each with 80 GB of GPU memory.
The code implementation is largely inspired by~\citet{zhang2026unleashing}, and the detailed LLM fine-tuning protocol is based on LLaMA-Factory~\citep{zheng2024llamafactory}.

\subsection{Item identifiers of \method}\label{subapp:ourtermid}

\method can be coupled with both term IDs and semantic IDs. 
We now elaborate on how we obtain the respective identifiers used as item identifiers.

\noindent\textbf{Term IDs (TIDs).}
We construct \textit{four keywords} for each item, where each keyword captures a different level of the item’s hierarchical characteristics. 
Specifically, the first keyword represents the most general attribute, shared by many items in the item space, while the last keyword represents the most specific attribute, shared by only a few items. 
To obtain these keywords, we first retrieve the 500 nearest neighbors of each item by sorting other items according to the cosine similarity between their textual-description representations and that of the target item.
Then, we build the following four item groups:
\begin{itemize}[leftmargin=*]
    \item \textbf{Group 1.} Three items randomly sampled from ranks 51 to 500.
    \item \textbf{Group 2.} Three items randomly sampled from ranks 11 to 50.
    \item \textbf{Group 3.} Three items randomly sampled from ranks 5 to 10.
    \item \textbf{Group 4.} The single most similar item.
\end{itemize}
Then, we prompt the LLM to generate the characteristics shared between the target item and the items in each group. 
The characteristic identified for Group 1 is used as the first keyword, while the characteristic identified for Group 4 is used as the last keyword.

After generating the four keywords for each item, we refine them by merging semantically equivalent keywords expressed in different forms, such as singular/plural variants and synonyms. 
This standardization improves consistency across items, making it easier for the LLM-based generative recommender to memorize and generate item representations.
To this end, for each keyword level, we aggregate the generated keywords and use an LLM to construct a mapping function that maps semantically equivalent terms to a unified keyword. 
For scalability, we process keywords in batches rather than feeding all terms to the LLM at once, where each batch contains a subset of the total keywords. 
We repeat this re-mapping process until convergence, defined as fewer than 10 keywords being changed between iterations.

\noindent\textbf{Semantic IDs (SIDs).}
For semantic item identifiers, we adopt the identifiers proposed by~\citet{he2026plum}.
Specifically, we apply RQ-VAE-based vector quantization~\cite{lee2022autoregressive} together with co-purchase-based contrastive learning across items.
We fix the three-level codebook sizes to $(1024,512,256)$, and set the training epochs, learning rate, weight decay, and contrastive-learning loss weight to $150$, $10^{-3}$, $10^{-6}$, and $10^{-1}$, respectively.

After obtaining the semantic identifier for each item, we perform continued pre-training to align the newly added special tokens corresponding to semantic identifier codes after vocabulary expansion.
To this end, we freeze the LLM-based generative recommender and its existing language-token embeddings, updating only the special tokens corresponding to the semantic identifier codes. 
Each training sample is item-level: the input is the item title, and the output is the semantic identifier sequence of the corresponding item.
This step is performed for 5 epochs with a learning rate of $10^{-3}$.

\subsection{Details for \method}\label{subapp:methoddetail}

\noindent\textbf{Backbone models and implementation.}
For the pretrained language embedding model, we use the Qwen-3-8B embedding model~\cite{zhang2025qwen3}. 
To build term identifiers, we use the instruction-tuned Qwen-3.5-9B model~\cite{yang2025qwen3} for draft keyword generation and keyword refinement. 
To build semantic identifiers, we use two-layer MLPs as encoders and decoders, with a hidden dimension of $256$.
All models are trained with the AdamW optimizer~\cite{loshchilov2017decoupled}, and LLM fine-tuning is implemented based on deepspeed~\cite{rasley2020deepspeed}.
For efficiency, model training is performed using bfloat16 precision, except for the 9B models, where we use float32 precision because bfloat16 led to numerical instability and NaN values.

\noindent\textbf{Hyperparameters.}
We train \method for 3 epochs with a fixed batch size of 128.
Gradient accumulation is used because processing all 128 data points in a single update is not feasible on 40 GB GPUs.
The co-purchased neighbor generation loss weight $\lambda_{1}$ is fixed to $1$, and the number of semantic neighbors $N_{2}$ is fixed to $5$.
We fix the window size in co-purchase neighbor generation $W$ to $5$.
We tune the following hyperparameters: the learning rate $\gamma$ in $\{5\times 10^{-5}, 10^{-4}\}$, 
the weight decay $\omega$ in $\{0, 10^{-2}\}$, 
the semantic neighbor generation loss weights $\lambda_{2}$ in $\{10^{-1}, 1\}$, 
and the numbers of collaborative neighbors $N_{1}$ in $\{5, 7\}$.

\begin{table}[t]
\centering
\caption{Hyperparameter settings for SID and TID in each dataset.}
\label{tab:hyperparameter_settings}
\resizebox{\linewidth}{!}{
\begin{tabular}{lcccccc}
\toprule
\multirow{2}{*}{} 
& \multicolumn{2}{c}{\textbf{Sports}} 
& \multicolumn{2}{c}{\textbf{Toys}} 
& \multicolumn{2}{c}{\textbf{Beauty}} \\
\cmidrule(lr){2-3} \cmidrule(lr){4-5} \cmidrule(lr){6-7}
& \textbf{SID} & \textbf{TID} 
& \textbf{SID} & \textbf{TID} 
& \textbf{SID} & \textbf{TID} \\
\midrule
$\gamma$ & $10^{-4}$ & $10^{-4}$ & $10^{-4}$ & $5\times10^{-5}$ & $10^{-4}$ & $5\times10^{-5}$ \\
$\omega$ & $10^{-2}$ & $0$ & $0$ & $0$ & $0$ & $0$ \\
$N_{1}$ & $5$ & $7$ & $5$ & $5$ & $5$ & $5$ \\
$\lambda_{2}$ & $1.0$ & $0.1$ & $1.0$ & $1.0$ & $1.0$ & $1.0$ \\
\bottomrule
\end{tabular}
}
\end{table}


\noindent\textbf{Inference of \method.}
For final inference, we use the standard beam search used in language models~\cite{sutskever2014sequence}.
Specifically, for each user, the LLM autoregressively generates 20 beams with a fixed number of tokens, and each beam is mapped to an item based on its identifier and the generated output.
For the final matching, we follow the matching strategy proposed by~\citet{zhang2026unleashing}.
Note that more than at least 99\% of beams are accurately matched to the existing item, which we elaborate in Appendix~\ref{subapp:outputvalidity}.
When an output beam corresponds to collided items (i.e., items sharing the same identifier), we select the most popular item as the final recommendation for that beam result.

\subsection{Baseline details}\label{subapp:baselines}

We elaborate on the details regarding the baseline methods used in this work.

\noindent\textbf{Traditional recommender systems.}
All models are trained with the AdamW~\cite{loshchilov2017decoupled} optimizer.
The backbone architectures of these methods follow those used in the original works.
We set the hidden dimension size and weight decay of all baseline methods to 128 and $10^{-6}$, respectively.
We train each model for 200 epochs and select the model checkpoint with the best validation Recall@10 results.
The learning rates of these models are tuned in $\{10^{-3}, 5\times10^{-4}, 10^{-4}\}$.

\noindent\textbf{Non-LLM-based GR models.}
All models are trained with the AdamW~\cite{loshchilov2017decoupled} optimizer.
For non-LLM-based GR models, we use the Transformer encoder-decoder architecture~\cite{vaswani2017attention} as the backbone generative retrieval model.
We set the hidden dimension size and weight decay of all baseline methods to 256 and $10^{-6}$, respectively.
We train each model for 100 epochs, and select the model checkpoint with the best validation Recall@10 results.
We set the output beam size to $20$, and tune the learning rate in $\{10^{-3}, 5\times10^{-4}, 10^{-4}\}$.

\noindent\textbf{LLM-based GR models.} 
All models are trained with the AdamW~\cite{loshchilov2017decoupled} optimizer.
For LLM-based GR models that use specialized identifiers and provide official implementations (i.e., GRAM~\cite{lee2025gram} and GRLM~\cite{zhang2026unleashing}), we use their implementations for term identifier generation. 
For methods without official implementations (i.e., AgenticTagger~\cite{xie2026agentictagger} and PLUM~\cite{he2026plum}), we implement the identifier generation step following the details provided in their manuscripts. 
We follow each baseline’s original training objective. 
When an external LLM is used to construct method-specific supervision or item representations, we use Gemini-2.5-Flash~\cite{comanici2025gemini} and otherwise keep the baseline pipeline consistent with its original formulation.
The overall training pipeline and hyperparameter configuration are kept the same as those of \method, except that we tune the learning rate for each method in $\{10^{-4}, 5\times 10^{-5}\}$.
All LLM-based GR methods use the same final inference strategy as in \method.

    \section{Additional experimental results}
    \label{app:addresult}

In this section, we provide additional experimental results omitted from the main manuscript due to space constraints.

\subsection{Informativeness of proposed neighbors}\label{subapp:data_analysis}
Recall that in Section~\ref{sec:method}, we introduce two types of neighbors that we use as our training objectives: collaborative neighbors and semantic neighbors.
In this section, we analyze the informativeness of these neighbors, particularly whether they provide effective signals for recommending items to users who do not benefit from the one-hop memorization.
To this end, we conduct three analyses: 
(RQ1) to what extent the neighbors of items in a user’s interaction sequence contain the target item, 
(RQ2) to what extent these neighbors benefit users who do not benefit from one-hop memorization, and 
(RQ3) whether these neighbors provide signals distinct from one-hop memorization.

\noindent\textbf{Setup.}
We consider three heuristic candidate sets:
\begin{itemize}[leftmargin=*]
    \item \textbf{Memorization-based:} Each user's candidates are the one-hop memorized items derived from the user's input sequence, following Section~\ref{subsec:analysis-setup}.
    \item \textbf{Collaborative-neighbor-based:} Each user's candidates are the top-15 most frequent collaborative neighbors of each item in the user's input sequence.
    \item \textbf{Semantic-neighbor-based:} Each user's candidates are the top-15 items with the most similar textual-description embeddings to each item in the user's input sequence.
\end{itemize}
Then, we apply the same candidate selection strategy as in \S~\ref{subsec:analysis-setup}, using the corresponding neighbor type.
Lastly, to answer these research questions, we measure: 
(A1) the number of users whose target item is included in each candidate set, 
(A2) the number of users who benefit from the two neighbor types but not from one-hop memorization, and 
(A3) the average Jaccard similarity between different candidate sets for each user.

\begin{figure*}[t!]
    \centering
    \includegraphics[width=0.8\linewidth]{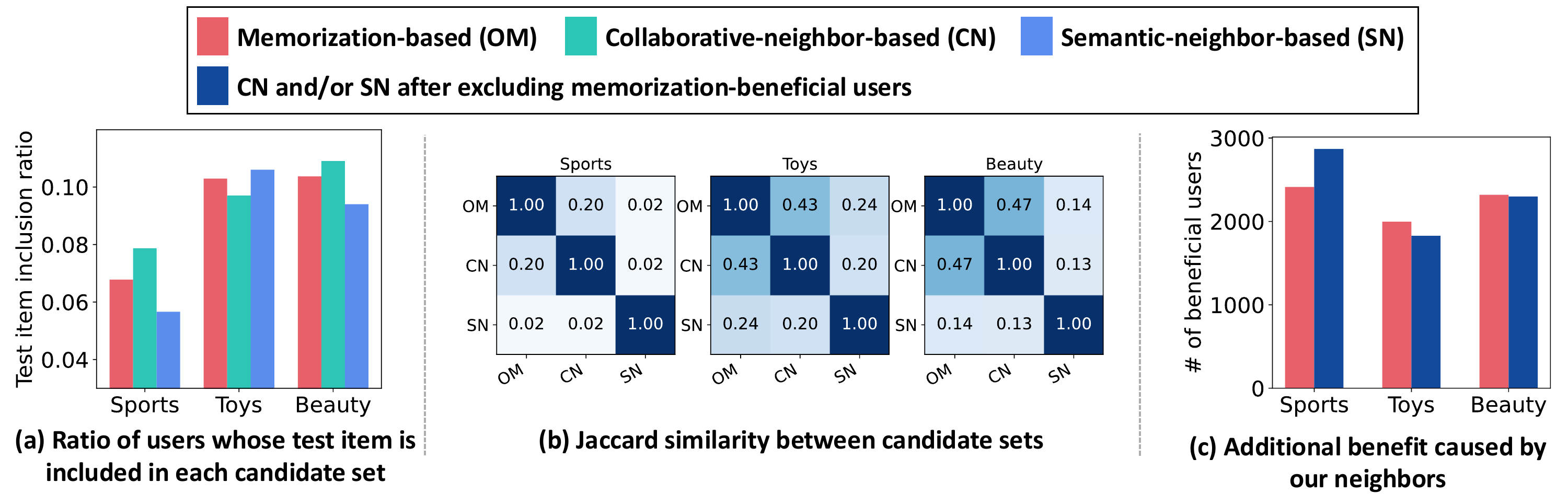}
    \caption{\textbf{Informativeness of our proposed neighbors.}
    The ratio of users benefiting from our collaborative and semantic neighbors is comparable to that of users benefiting from one-hop memorization (see (a)). 
    Moreover, the three neighbor types have a Jaccard similarity below 0.5, suggesting that they are not merely redundant (see (b)).
    In addition, among non-memorization-benefiting users, our neighbors benefit a group of users comparable in size to the memorization-benefiting group (see (c)).
    }
    \label{fig:informativeness}
\end{figure*}

\noindent\textbf{Result.}
Results are provided in Figure~\ref{fig:informativeness}.
As shown in (a), our collaborative and semantic neighbors cover a fraction of users comparable to one-hop memorization. 
In addition, as shown in (b), their pairwise Jaccard similarity with one-hop neighbors remains below 0.5, indicating that these neighbor types provide non-redundant information. 
Lastly, as shown in (c), among non-memorization-benefiting users, our neighbors cover a group comparable in size to the memorization-benefiting group, suggesting that our supervision signals provide predictive information comparable to or even stronger than one-hop memorization.

\begin{figure}[t]
    \centering
    \includegraphics[width=\linewidth]{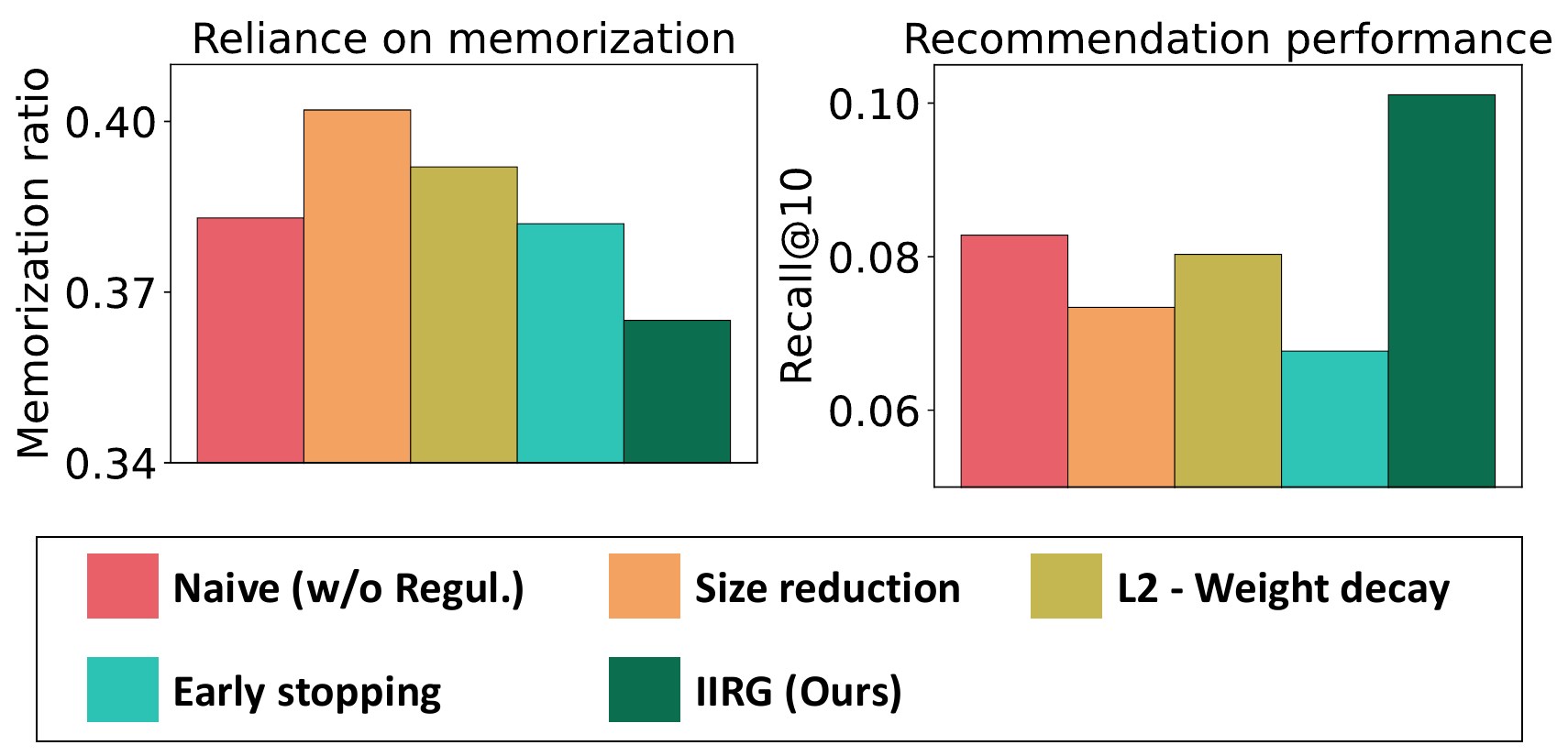}
    \caption{\textbf{Effect of existing regularization techniques.}
    In the Beauty dataset, existing regularization techniques cannot effectively reduce the LLM's one-hop memorization behavior and/or hurt the LLM's recommendation performance.
    }
    \label{fig:regularization}
\end{figure}

\subsection{Effect of regularization techniques}\label{subapp:regularization}

In this section, we explore whether common regularization techniques can alleviate the LLM's reliance on one-hop transitions.

\noindent\textbf{Setup.}
We use four regularization techniques in our analysis: 
(1) LoRA~\cite{hu2022lora}; 
(2) model-size reduction, using a 0.8B-parameter LLM while all other LLMs in our analysis use 4B-parameter models; 
(3) $L_{2}$ weight decay in the optimizer; and 
(4) early stopping at epoch 2 instead of the original 3 epochs.
We use term IDs as item identifiers.
We compare the methods by fine-tuning the LLM for the next-item prediction task with each regularization technique.
We use the Beauty dataset.

\noindent\textbf{Result 1: Failure of LoRA.}
We find that LoRA fails to generate valid recommendation outputs in our setting.
Specifically, 94\% of the outputs generated by the LoRA-fine-tuned LLM are invalid.
These invalid outputs either consist of natural-language reasoning that does not explicitly mention an item (e.g., ``Given the user interaction history, the user is likely to purchase camping products''), or contain item identifiers that do not exist in the item space.
This trend holds across different LoRA ranks: $r=16$, corresponding to 1\% of the total parameters, $r=64$, corresponding to 7\% of the total parameters, and $r=256$, corresponding to 13\% of the total parameters.
This result suggests that training a sufficient number of LLM parameters is necessary for enabling the LLM to both follow the desired output format and memorize valid item identifiers.

\noindent\textbf{Result 2: Other methods.}
As shown in Figure~\ref{fig:regularization}, all three regularization techniques fail to reduce one-hop memorization without hurting recommendation performance. 
In contrast, \method reduces memorization behavior while improving downstream task performance, achieving the largest gains in both aspects. 
This suggests that \method addresses the memorization behavior more effectively than common regularization techniques.






\subsection{Discussion on excluding one-hop transitions from collaborative neighbors}\label{subapp:onehopout}

Note that when constructing collaborative neighbors (\S~\ref{subsec:method-detail}), we treat all items appearing within a window of size $W$ around each item as its collaborative neighbor candidates. 
This often includes the one-hop successor as well, which may preserve the model’s reliance on one-hop memorization. 
We further analyze whether removing such successors leads to performance improvements.

\begin{table}[t]
\centering
\caption{\textbf{Exclusion of one-hop transitions for collaborative neighbors}.
The current design choice of collaborative neighbors for \method outperforms the alternative that excludes one-hop transition items (OT).
}
\label{tab:one-hopremoval}
\resizebox{0.8\columnwidth}{!}{
\begin{tabular}{l | cc | cc}
\toprule
\multirow{2}{*}{} 
& \multicolumn{2}{c|}{Toys} 
& \multicolumn{2}{c}{Beauty} \\
\cmidrule(lr){2-3} \cmidrule(lr){4-5}
& R@10 & N@10 & R@10 & N@10 \\
\midrule
w/o OT & 0.1022 & 0.0605 & 0.0961 & 0.0569 \\
w/ OT (\method) & \textbf{0.1097} & \textbf{0.0642} & \textbf{0.1011} & \textbf{0.0606} \\
\bottomrule
\end{tabular}
}
\end{table}

\noindent\textbf{Setup.}
An alternative design is to exclude the one-hop successor of the corresponding item and use only the remaining $2W-1$ items as co-purchase neighbors, which more explicitly reduces reliance on one-hop memorization.
We then apply the same selection strategy as in~\S~\ref{subsec:method-detail} to select $N_1$ items, and use the resulting neighbors as supervision signals for the collaborative neighbor generation task. 
We use TIDs as item identifiers.

\noindent\textbf{Result.}
As shown in Table~\ref{tab:one-hopremoval}, our current design outperforms the alternative that explicitly removes one-hop transition items. 
We hypothesize that this removal may bias the LLM too strongly toward multi-hop relations, causing it to underutilize one-hop relations that still provide useful predictive signals for certain user groups.
Therefore, the current design of \method{} may better balance one-hop and multi-hop relations, which are both important for accuracy.

\begin{table}[t]
    \centering
    \caption{\textbf{Additional ablation study 1.} 
    In the Toys dataset, \method outperforms the random-neighbor-based variant, demonstrating that training based on predictive neighbors is crucial for strong performance.}
    \label{tab:ablation1}
    \begin{tabular}{c|ccc}
        \toprule
        & \textbf{R@5} & \textbf{NDCG@5} \\
        \midrule
        Naive & 0.0634 & 0.0438 \\
        Random & 0.0539 & 0.0361 \\
        \midrule
        \method & \textbf{0.0780} & \textbf{0.0541} \\
        \bottomrule
    \end{tabular}
\end{table}

\begin{table}[t]
    \centering
    \caption{\textbf{Additional ablation study 2.} 
    \method outperforms our neighbor-based heuristics, demonstrating that training LLMs to learn such relations is more effective than simply using those relations directly.
    Recall@10 of each model is reported.}
    \label{tab:ablation2}
    \begin{tabular}{c|ccc}
        \toprule
        & \textbf{Sports} & \textbf{Toys} & \textbf{Beauty} \\
        \midrule
        {V1} & 0.0467 & 0.0769 & 0.0671 \\
        {V2} & 0.0464 & 0.0836 & 0.0742 \\
        {V3} & 0.0294 & 0.0636 & 0.0496 \\
        {V4} & 0.0321 & 0.0810 & 0.0615 \\
        \midrule
        \method & \textbf{0.0585} & \textbf{0.1097} & \textbf{0.1011} \\
        \bottomrule
    \end{tabular}
\end{table}

\subsection{Additional ablation studies}\label{subapp:additionalablation}

We conduct two ablation studies to analyze the design choices of \method. 
Specifically, we consider: (1) an LLM trained with randomly sampled neighbors instead of the collaborative and semantic neighbors introduced in \S~\ref{sec:method}; 
and (2) heuristic methods that directly use collaborative or semantic neighbors as recommendation outputs.

\noindent\textbf{Setup 1.}
We consider a variant that randomly samples collaborative and semantic neighbors instead of the strategies described in \S~\ref{subsec:method-detail}, which we call Random.
We also examine whether this variant harms the performance of Naive, which is trained solely on next-item prediction.
We use TIDs as item IDs.

\noindent\textbf{Result 1.}
As shown in Table~\ref{tab:ablation1}, (1) \method{} outperforms Random, and (2) Random even degrades the performance of Naive. 
This suggests that, for \method{} to achieve strong performance, its supervision signal should be predictive of users' test items.

\noindent\textbf{Setup 2.} 
We consider four heuristics:
(V1) For each item in the user’s input sequence, we sample the five most frequently co-purchased items, aggregate the candidates, and select 10 final recommendations based on their multiplicity, with popularity-based tie-breaking.
(V2) Overall structure is the same as V1, but we only aggregate candidates of the two most recently interacted items of the user.
(V3) For each item in the user’s input sequence, we sample the top five items with highest textual representation similarity, aggregate the candidates, and select 10 final recommendations based on their multiplicity.
(V4) Overall structure is the same as V3, but we only aggregate candidates of the last-2 interacted items of the user.

\noindent\textbf{Result 2.}
As shown in Table~\ref{tab:ablation2}, \method outperforms all variants, demonstrating that training LLMs to capture neighbor relations for recommendation is more effective than simply using our collaborative and semantic relations directly.





\begin{table}[t]
\centering
\caption{\textbf{Results on other domains.}
\method also outperforms the Naive baseline on both the ML-1M and Yelp datasets.
}
\label{tab:otherdomainresult}
\begin{tabular}{l | cc | cc}
\toprule
& \multicolumn{2}{c|}{Yelp}
& \multicolumn{2}{c}{ML-1M} \\
& R@5 & N@5 & R@5 & N@5 \\
\midrule
Naive
& 0.0490 & 0.0335
& 0.1005 & 0.0654 \\
\method{}
& \textbf{0.0496} & \textbf{0.0341}
& \textbf{0.1464} & \textbf{0.0988} \\
\bottomrule
\end{tabular}
\end{table}

\subsection{Generalization across diverse domains}\label{subapp:yelpresult}

We evaluate the effectiveness of \method{} in recommendation beyond the Amazon review domains.

\noindent\textbf{Setup.}
We evaluate \method on two additional domains: restaurant reviews using Yelp\footnote{\url{https://www.yelp.com/dataset}} and movie ratings using MovieLens-1M~\cite{harper2015movielens} (ML-1M).
We follow the same training and evaluation protocol as in \S~\ref{subsec:mainperformance} and use TIDs as item identifiers.
On both datasets, we examine whether \method outperforms Naive, which trains the LLM solely with the next-item prediction objective.

\noindent\textbf{Results.}
As shown in Table~\ref{tab:otherdomainresult}, \method{} consistently outperforms Naive on both datasets, demonstrating that its effectiveness extends beyond the Amazon review datasets.

\subsection{Discussion on joint training}\label{subapp:whyjoint}
Our preliminary analysis shows that joint learning outperforms continued pre-training, where the LLM is first trained on our two neighbor generation tasks and then further fine-tuned for next-item prediction.
We hypothesize that this is because, in the latter approach, the LLM tends to forget the rich item relations learned through our tasks and instead reverts to focusing on local transition patterns learned from the next-item prediction task, consistent with observations in LLM continued fine-tuning literature~\cite{luo2025empirical}.

\subsection{Efficient semantic neighbor generation}\label{subapp:efficient_semantic}

\begin{figure}[t]
    \centering
    \includegraphics[width=\linewidth]{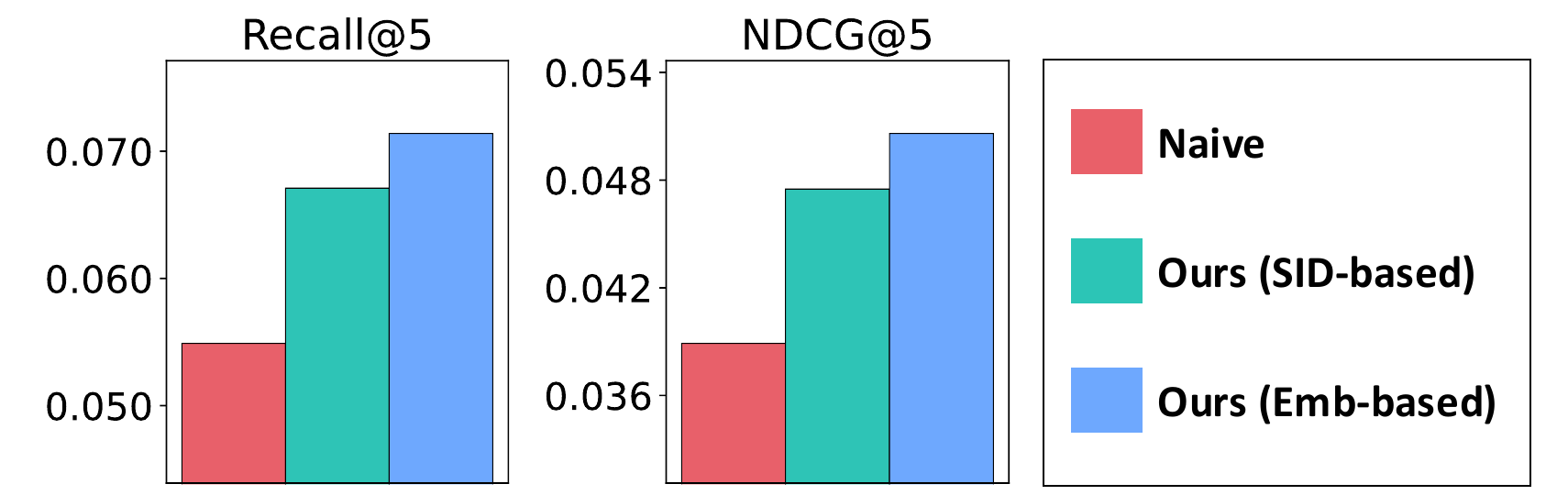}
    \caption{\textbf{Performance of diverse semantic neighbor search methods for \method.}
    Searching semantic neighbors with SIDs still improves the recommendation performance of the LLM trained solely with next-item prediction (Naive), while the performance drop compared with full dense embedding search (Emb) remains relatively small.
    }
    \label{fig:efficientsearch}
\end{figure}

\noindent\textbf{Search strategy.}
Note that semantic neighbors are obtained from item embeddings encoded from item textual descriptions (Section~\ref{subsec:method-detail}). 
A full search over all item embeddings incurs quadratic cost with respect to the number of items. 
To reduce this, we introduce an efficient search strategy based on each item’s semantic identifier. 
Instead of performing dense search over all item embeddings, we select the top-$K$ items according to the number of semantic identifier codes they share with the target item. 
Ties are broken by prioritizing more popular items.

\noindent\textbf{Setup.}
We compare the performance of the two retrieval strategies, as well as the performance without our training strategy. 
In all cases, we use semantic identifiers as item identifiers.
We use the Toys dataset for training and evaluation.

\noindent\textbf{Result.}
As shown in Figure~\ref{fig:efficientsearch}, even when semantic neighbors are retrieved using semantic IDs (SIDs) rather than dense embeddings, the resulting model achieves clear gains over the Naive LLM trained only with next-item prediction. 
Moreover, the gap from full dense embedding search (Emb) remains small, suggesting that SID-based retrieval offers an efficient alternative strategy with limited performance loss.

\begin{table}[t]
\centering
\caption{\textbf{Output validity}. 
We report the percentage of generated beams that are successfully mapped to the item space.
In all the cases, \method achieves at least 99\% of output validity.}
\label{tab:output_validity}
\begin{tabular}{lccc}
\toprule
\textbf{ID Type} & \textbf{Sports} & \textbf{Toys} & \textbf{Beauty} \\
\midrule
SIDs & 100\% & 99\% & 100\% \\
TIDs & 99\% & 99\% & 99\% \\
\bottomrule
\end{tabular}
\end{table}


\subsection{Output validity analysis}\label{subapp:outputvalidity}
One undesirable behavior in generative recommendation is generating non-existent items. 
We therefore verify whether \method improves valid item generation for both semantic-ID and term-ID settings.

\noindent\textbf{Setup.}
Recall that we generate $20$ beams for each user. 
Across all beams for all users, we compute the ratio of generated beams that are successfully decoded and mapped to the item space.

\noindent\textbf{Result.}
As shown in Table~\ref{tab:output_validity}, around 99\% of the outputs generated by \method{} can be mapped to actual items, demonstrating that \method rarely produces hallucinated item identifiers.

\begin{figure*}[t]
    \centering
    \includegraphics[width=\linewidth]{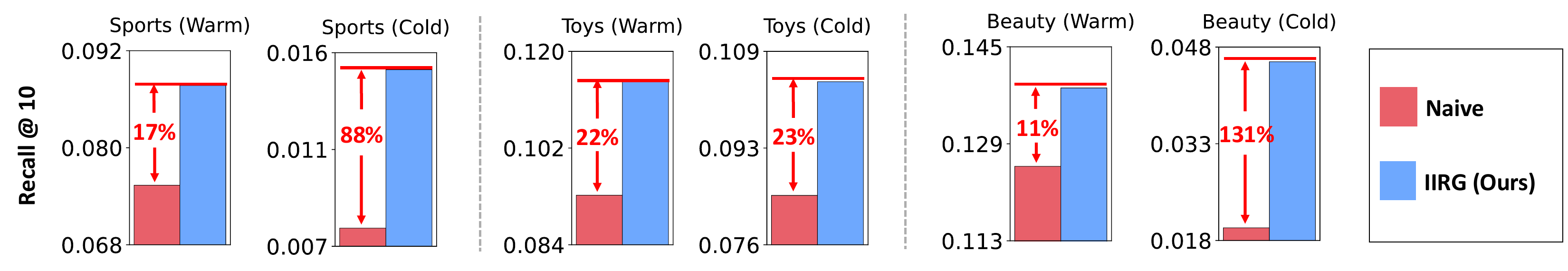}
    \caption{\textbf{In-depth analysis regarding warm- and cold-start items.}
    The gain of \method over the LLM trained solely with next-item prediction (Naive) is consistently larger for users whose test items have limited user interactions than for users whose test items have rich user interactions.
    }
    \label{fig:colditem}
\end{figure*}

\begin{figure*}[t]
    \centering
    \includegraphics[width=\linewidth]{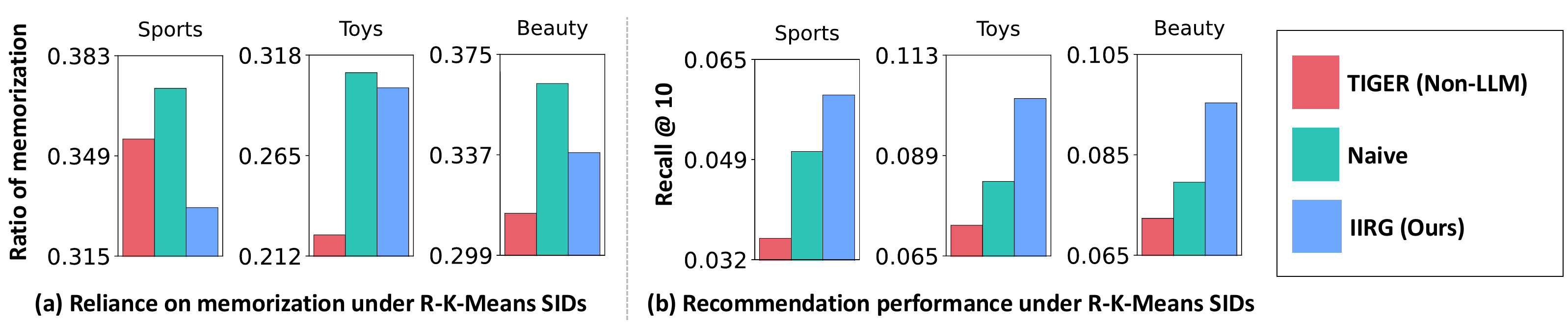}
    \caption{\textbf{Generalization to other semantic IDs (SIDs).}
    The observation that LLMs rely more heavily on one-hop memorization than TIGER, as well as the effectiveness of \method{} in reducing memorization and improving recommendation performance, also holds under Residual-K-Means (R-K-Means) SIDs.
    }
    \label{fig:rkmeans}
\end{figure*}

\begin{figure}[t]
    \centering
    \includegraphics[width=\linewidth]{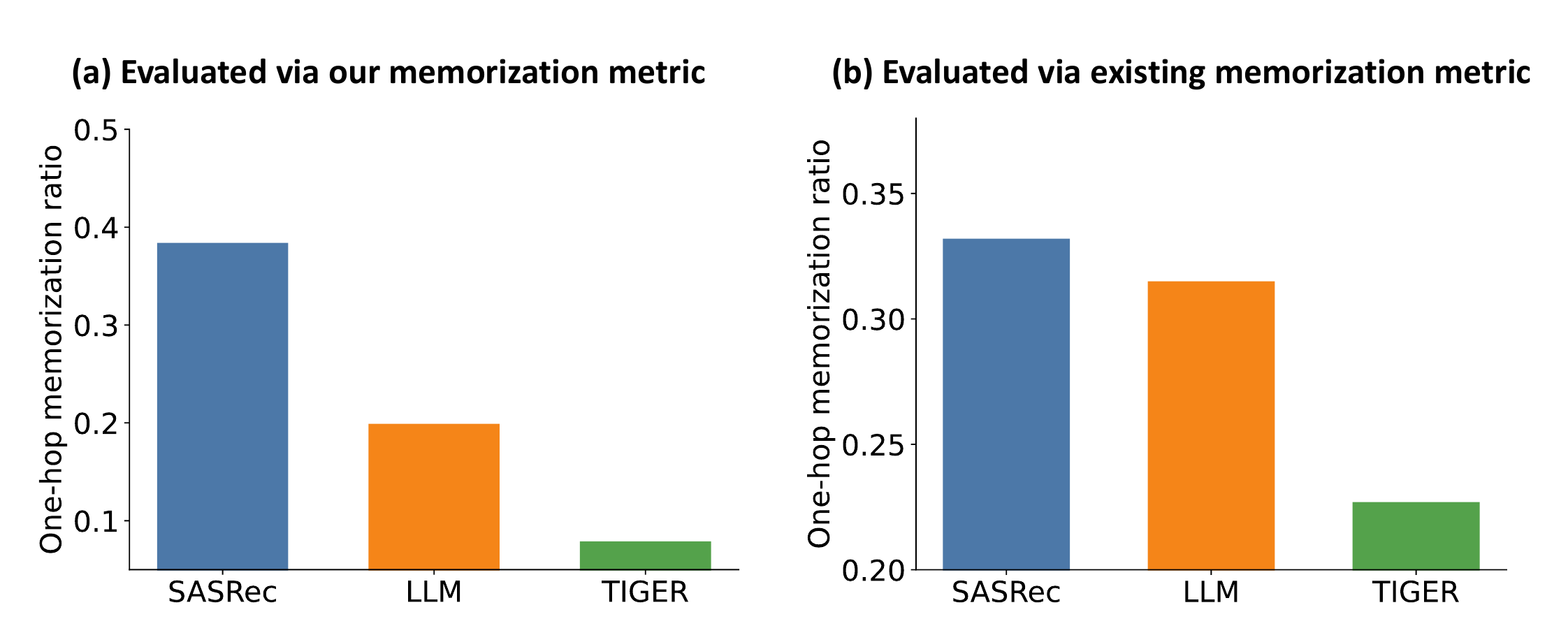}
    \caption{\textbf{Unified one-hop memorization ratio analysis across diverse metrics.}
    We observe the same trend using both our metric and the metric proposed by \citet{ding2026well}.
    Specifically, ID-based model (SASRec) shows the highest one-hop memorization, followed by LLM and TIGER.
    }
    \label{fig:unifiedmemorization}
\end{figure}

\begin{figure}[t]
    \centering
    \includegraphics[width=\linewidth]{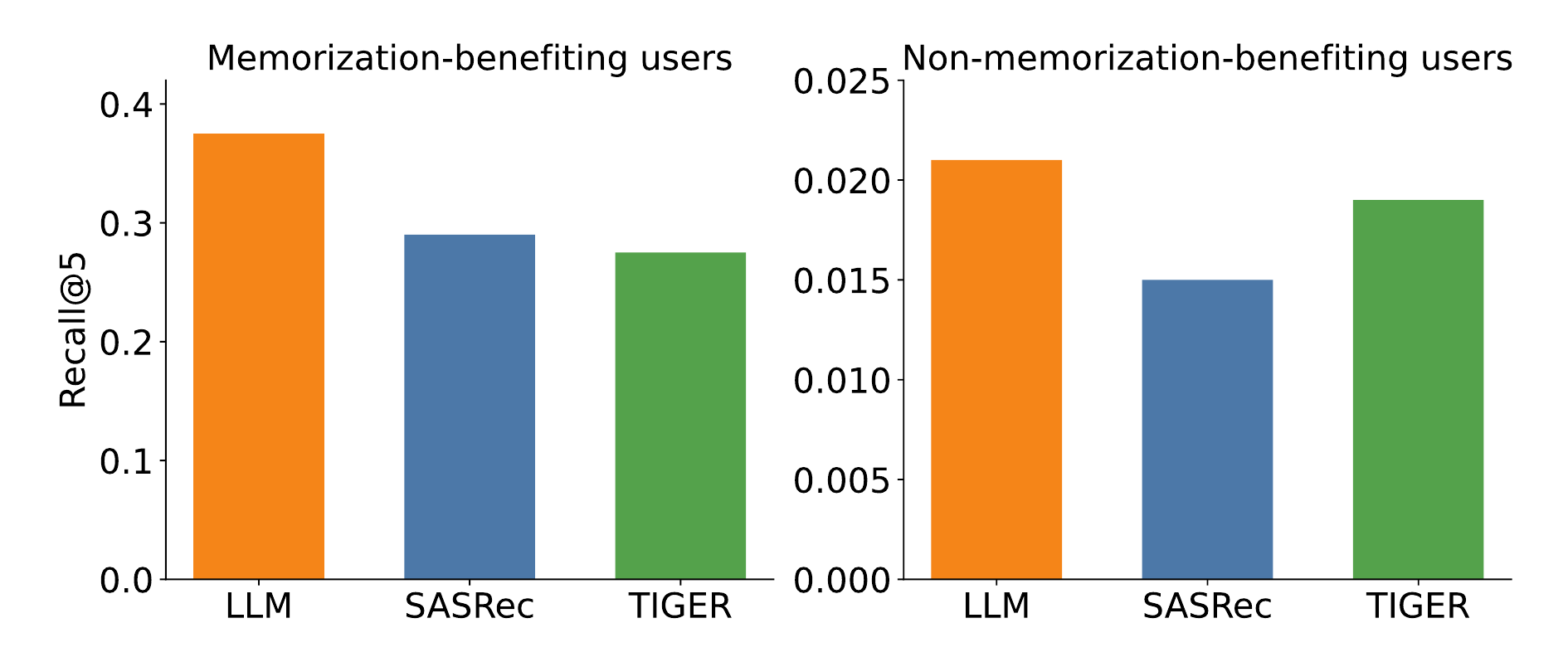}
    \caption{\textbf{Unified performance analysis.}
    The LLM shows the highest performance in either user group. 
    ID-based method outperforms TIGER in memorization-benefiting users, while the trend is reversed in non-memorization-benefiting users.
    }
    \label{fig:unifiedperformance}
\end{figure}

\subsection{Cold-start-item recommendation}\label{subapp:coldstart}

We examine how the performance improvement of \method over Naive, an LLM trained solely with next-item prediction, varies between cold-start and warm-start target items.
Specifically, we classify items with low purchase frequencies as cold-start items and those with high purchase frequencies as warm-start items.

\noindent\textbf{Setup.}
We divide users into two groups based on the interaction frequency of their target item: (1) \textit{cold-item users}, whose target item has at most $8$ interactions, and (2) \textit{warm-item users}, whose target item has more than $8$ interactions. 
We then compare the performance improvement within each user group.

\noindent\textbf{Result.}
As shown in Figure~\ref{fig:colditem}, the percentage improvement of \method{} over Naive is larger for cold-item users than for warm-item users, suggesting that \method{} is especially effective at recommending cold-start items. 
We hypothesize that this is because \method{}’s neighbor generation tasks, particularly semantic neighbor generation, provide additional supervision for cold-start items, leading to larger performance gains.






\subsection{Generalization to other semantic IDs}\label{subapp:rkmeans}

As detailed in Section~\ref{sec:experiment}, our main analysis and experiments use PLUM~\cite{he2026plum} SIDs.
We further examine whether our findings generalize to other types of SIDs.

\noindent\textbf{Setup.}
We use Residual K-Means (R-K-Means) SIDs~\cite{zhou2025openonerec} as item identifiers.
We then analyze the memorization reliance and recommendation performance of Naive, an LLM trained solely with next-item prediction, and \method, both using R-K-Means SIDs.

\noindent\textbf{Result.}
As shown in Figure~\ref{fig:rkmeans}, Naive relies more heavily on memorization than TIGER, a non-LLM-based GR model.
Moreover, \method reduces Naive's memorization behavior and improves its recommendation performance across all datasets.
These results suggest that our observations and the effectiveness of \method are not restricted to a particular SID design.

\subsection{Unified memorization analysis}\label{subapp:unifiedanalysis}

As discussed in Section~\ref{sec:relatedwork},~\citet{ding2026well} also analyze memorization in non-LLM-based generative recommenders in comparison with ID-based recommender systems.
In this section, we explore the unified analysis that considers all memorization metrics and recommender systems investigated in either ours or~\citet{ding2026well}.

\noindent\textbf{Setup.}
For the definition of memorization, we use either our definition (Section~\ref{sec:analysis}) and that of~\citet{ding2026well}. 
Moreover, we use ID-based recommender system (SASRec~\citep{kang2018self}), non-LLM-based GR (TIGER~\citep{rajput2023recommender}), and LLM-based GR (Qwen3-1.7B).
We evaluate on the Toys dataset.

\noindent\textbf{Result.}
Our unified analysis confirms the findings of both our analysis (Section~\ref{sec:analysis}) and \citet{ding2026well}. 
Specifically, ID-based recommender systems exhibit the highest one-hop memorization, followed by LLM-based GR and non-LLM-based GR, as shown in Figure~\ref{fig:unifiedmemorization}. 
These conclusions hold across all memorization metrics considered.
Among memorization-benefiting users (defined in Section~\ref{sec:analysis}), LLM-based GR achieves the best recommendation performance, followed by ID-based recommender systems and non-LLM-based GR, as shown in Figure~\ref{fig:unifiedperformance}. 

\begin{table}[t]
\centering
\caption{\textbf{Recommendation performance}.
Compared to the three variants (i.e., neighbor selection (V1), pairwise generation (V2), and shuffled-list generation (V3)), \method consistently shows the best performance.}
\label{tab:alternativeresults}
\begin{tabular}{lcccc}
\toprule
\textbf{Metric} & \textbf{V1} & \textbf{V2} & \textbf{V3} & \textbf{\method} \\
\midrule
Recall@5 & 0.058 & 0.060 & 0.061 & 0.065 \\
NDCG@5   & 0.041 & 0.042 & 0.043 & 0.047 \\
\bottomrule
\end{tabular}
\end{table}

\subsection{Alternative training design}\label{subapp:alternativetraining}
Note that we formulate each collaborative neighbor generation task as a sequence generation task. 
In this section, we evaluate an alternative formulation.

\noindent\textbf{Setup.}
Instead of doing sequence generation task, we use a pairwise anchor-neighbor generative task.
Specifically, we consider two tasks. 
The first is a prediction task (V1), in which the LLM selects the ground-truth neighbor of an anchor item from a candidate set containing that neighbor and randomly sampled items. 
The second is a generation task (V2), in which the LLM generates a single neighbor item given an anchor item.
We then evaluate the effectiveness of each method when its corresponding task is used as an auxiliary training objective.
Moreover, for the sequence generation task, we include a baseline in which neighbors are randomly shuffled rather than ordered by relevance (V3).
We conduct the evaluation on the Toys dataset using Qwen3-1.7B as the backbone LLM.

\noindent\textbf{Result.}
As shown in Table~\ref{tab:alternativeresults}, \method consistently outperforms these variants. 
This result supports that the effectiveness of our design over the alternatives.

\begin{figure*}[t]
    \centering
    \includegraphics[width=\linewidth]{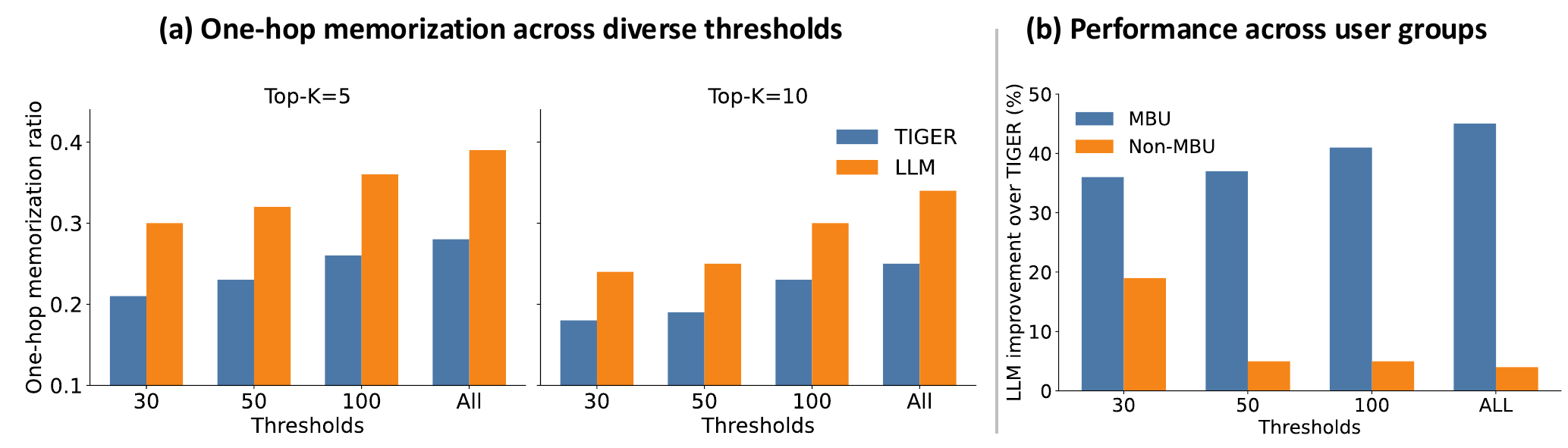}
    \caption{\textbf{Robustness against the cutoff thresholds.}
    Our key finding holds across diverse thresholds.
    }
    \label{fig:thresholds}
\end{figure*}

\subsection{Diverse cutoff for analysis}\label{subapp:cutoff}

Note that we use cutoff-threshold of $50$ in our analysis in Section~\ref{sec:analysis}.
In this section, we evaluate whether our finding holds in diverse cutoff thresholds as well.

\noindent\textbf{Setup.}
Beyond the cutoff of $50$, we use three more cutoff thresholds: $30$, $100$, and without cutoff (i.e., using all the one-hop successors as one-hop candidates).
Then, we fix the rest of the setting same as in Section~\ref{sec:analysis}.
We conduct the evaluation on the Toys dataset using Qwen3-1.7B as the backbone LLM.

\noindent\textbf{Result.}
As shown in Figure~\ref{fig:thresholds}, our findings remain consistent with those in Section~\ref{subsec:analysis-result} across a wide range of cutoff thresholds. 
Specifically, LLM-based GR exhibits higher one-hop memorization than TIGER, a non-LLM-based GR, and its performance gains over TIGER are concentrated among memorization-benefiting users. 
Across these thresholds, our method reduces one-hop memorization in LLM-based GR and yields larger relative performance gains for non-memorization-benefiting users than for memorization-benefiting users.


    \section{Discussions on memorization}
    \label{app:memorizationdiscussion}

In this section, we clarify several aspects of our work that may otherwise be misunderstood.

\subsection{Clarification on implications and our goal}\label{subapp:clarifyingimplication}
First, importantly, we clarify that one-hop memorization is not inherently undesirable, as observed one-hop transitions provide useful recommendation signals (see \S~\ref{subsec:analysis-result}). 
However, these signals have limited coverage and provide little information for users whose target items are not connected to their histories through training-time one-hop transitions. 
Our goal is therefore not to eliminate one-hop memorization, but to complement it with richer collaborative and semantic item relations.

\subsection{Scope of one-hop memorization}
\label{subapp:memorizationscope}
In this work, we use \textit{one-hop memorization} as a behavior-level term for a model's reliance on one-hop item transitions observed during recommendation training.
Our metric measures how often the model recommends training-observed successors of items in a test user's interaction sequence.
It does not determine whether this behavior results from recalling specific training examples or from learning recurring transition patterns.
Accordingly, our analysis focuses on observable reliance on training-observed one-hop transitions, rather than the model's internal memorization mechanism.
Moreover, prior work compares generative recommendation with traditional ID-based recommendation by categorizing test instances according to whether their item transitions were observed during training~\cite{ding2026well}. 
In contrast, we focus specifically on generative recommendation and show that pretrained LLM-based GR exhibits stronger reliance on such transitions than a non-LLM GR baseline. 
We provide further details on the unified pipeline in Appendix~\ref{subapp:unifiedanalysis}.

\subsection{Distinction from existing auxiliary training methods for LLM-based GR}
\label{subapp:distinction}
We do not claim that collaborative relations, semantic relations, or multi-task learning are individually new concepts in recommendation.
IIRG’s novelty lies in connecting the behavioral diagnosis in \S~\ref{sec:analysis} to a distinct supervision design: explicit generation of multi-hop collaborative and semantic neighbors jointly with next-item prediction. 
Unlike existing auxiliary tasks centered on knowledge distillation, item-ID alignment, rating prediction, or sequence reconstruction, \method directly supervises item-to-item relations. 
We further validate whether this supervision addresses the specific user group identified by our analysis, as shown in \S~\ref{subsec:designgoal}.

\subsection{Experimental evidence on method design}
Our experiments further demonstrate that the proposed design is effective and that its gains cannot be attributed to a trivial adaptation, thereby supporting its methodological novelty.
Specifically, our additional analyses further distinguish \method from simply adding more training examples or directly recommending neighboring items  (see Appendix~\ref{subapp:additionalablation}).
Training with random neighbors does not yield improvements, and directly using collaborative or semantic neighbors as recommendations underperforms \method (see Appendix~\ref{subapp:additionalablation}).
These results suggest that \method's gains arise from learning the proposed item--item relations within the recommendation objective, rather than merely increasing the amount of training data.

\section{AI Assistant Usage}
\label{app:assistant}

We used large language models in two ways: (1) to assist with code implementation of the proposed method and baseline methods, and (2) to refine the writing of this paper.

\section{License}
\label{app:license}

Our proposed model and its implementation code are released under the MIT License. 
In addition, for the baseline methods, we use only open-source code implementations whose licenses permit their use in this work.

\end{document}